\documentclass{revtex4}
\usepackage{eurosym}
\usepackage{graphicx}
\usepackage{amsmath}
\usepackage{amsfonts}
\usepackage{amssymb}
\usepackage[caption=false]{subfig}
\usepackage{float}
\usepackage[running,mathlines,displaymath]{lineno}
\begin{document}
\title{Robust $\mathcal{PT}$ symmetry of two-dimensional fundamental and
	vortex solitons supported by spatially modulated nonlinearity}
\author{Eitam Luz$^1$, Vitaly Lutsky,$^1$ Er'el Granot,$^2$ and Boris A. Malomed$^{1,3}$}
\address{$^1$Department of Physical Electronics, School of Electrical Engineering,
Faculty of Engineering, Tel Aviv University, Tel Aviv 69978, Israel\\
$^2$Department of Electrical and Electronic Engineering, Ariel University,
Ariel, Israel\\
$^3$Center for Light-Matter Interaction, Tel Aviv University, Tel Aviv
69978, Israel}
\begin{abstract}
The real spectrum of bound states produced by $\mathcal{PT}$-symmetric
Hamiltonians usually suffers breakup at a critical value of the strength of
gain-loss terms, i.e., imaginary part of the complex potential. The breakup
essentially impedes the use of $\mathcal{PT}$-symmetric systems for various
applications. On the other hand, it is known that the $\mathcal{PT}$
symmetry can be made unbreakable in a one-dimensional (1D) model with
self-defocusing nonlinearity whose strength grows fast enough from the
center to periphery. The model is nonlinearizable, i.e., it does not have a
linear spectrum, while the (unbreakable) $\mathcal{PT}$ symmetry in it is
defined by spectra of continuous families of nonlinear self-trapped states
(solitons). Here we report results for a 2D nonlinearizable model whose $
\mathcal{PT}$ symmetry remains unbroken for arbitrarily large values of the
gain-loss coefficient. Further, we introduce an extended 2D model with the
imaginary part of potential $\sim xy$ in the Cartesian coordinates. The
latter model is not a $\mathcal{PT}$-symmetric one, but it also supports
continuous families of self-trapped states, thus suggesting an extension of
the concept of the $\mathcal{PT}$ symmetry. For both models, universal
analytical forms are found for nonlinearizable tails of the 2D modes, and
full exact solutions are produced for particular solitons, including ones
with the unbreakable $\mathcal{PT}$ symmetry, while generic soliton families
are found in a numerical form. The $\mathcal{PT}$-symmetric system gives
rise to generic families of stable single- and double-peak 2D solitons
(including higher-order radial states of the single-peak solitons), as well
as families of stable vortex solitons with $m=1$, $2$, and $3$. In the model
with imaginary potential $\sim xy$, families of single- and multi-peak
solitons and vortices are stable if the imaginary potential is subject to
spatial confinement. In an elliptically deformed version of the latter
model, an exact solution is found for vortex solitons with $m=1$.
\end{abstract}

\maketitle

\section*{Introduction}

While wave functions of quantum systems may be complex, spectra of their
energy eigenvalues must be real, which is usually secured by restricting the
underlying Hamiltonian to be Hermitian \cite{qm}. However, the condition of
the reality of the energy spectrum does not necessarily imply that it is
generated by an Hermitian Hamiltonian. Indeed, it is well known that
non-Hermitian Hamiltonians obeying the parity-time ($\mathcal{PT}$) symmetry
may also produce entirely real spectra~\cite%
{bender1,dorey,bender2,bender3,review,ptqm}. In terms of the single-particle
complex potential,
\begin{equation}
P(\mathbf{r})\equiv V(\mathbf{r})+iW(\mathbf{r}),  \label{U}
\end{equation}%
the $\mathcal{PT}$ symmetry requires its real and imaginary parts to be even
and odd functions of coordinates \cite{bender1}: $V(\mathbf{r})=V(-\mathbf{r}%
),W(-\mathbf{r})=-W(\mathbf{r})$,~i.e.,~
\begin{equation}
P(-\mathbf{r})=P^{\ast }(\mathbf{r}),  \label{minus}
\end{equation}%
where the asterisk stands for the complex conjugate. Actually, Hamiltonians
which keep $\mathcal{PT}$ symmetry may be transformed into Hermitian ones
\cite{Mostafazadeh,Barash2,Barash}.

In the general case, the energy spectrum generated by the $\mathcal{PT}$%
-symmetric potential remains real (physically relevant) below a certain
critical value of the strength of the imaginary part of the underlying
potential, $W(\mathbf{r})$ in Eq. (\ref{U}), which is a threshold of the $%
\mathcal{PT}$ symmetry breaking. Above the critical value, the system is
made unstable by emerging imaginary parts of energy eigenvalues. In some
models, the breakup of the $\mathcal{PT}$ symmetry may follow the onset of
the jamming anomaly, which means a transition from increase to decrease of
the power flux between the spatially separated gain and loss spots with the
growth of the gain-loss coefficient \cite{jamming1}. The fragility of the $%
\mathcal{PT}$ symmetry essentially limits the use of this property in
applications, where new effects, such as unidirectional transmissivity \cite%
{uni}, enhanced absorption of light \cite{Longhi}, lasing in microrings \cite%
{exp5}, acoustic sensors \cite{sensors}, as well as the operation of $%
\mathcal{PT}$-symmetric metamaterials \cite{exp4} and microcavities \cite%
{exp6} strengthen with the increase of the gain-loss coefficient.

Thus far, the $\mathcal{PT}$ symmetry was not experimentally realized in
quantum systems, and, moreover, it was argued that, strictly speaking, $%
\mathcal{PT}$-symmetric systems do not exist in the framework of the quantum
field theory \cite{Szameit}. On the other hand, a possibility to implement
the concept of the $\mathcal{PT}$ symmetry in terms of classical physics was
predicted for optical media with symmetrically placed gain and loss elements
\cite{theo1}-\cite{Kominis}, which is based on the similarity between the
Schr\"{o}dinger equation in quantum mechanics and the paraxial-propagation
equation for optical waveguides. Experimentally, this possibility was
implemented in several waveguiding settings \cite{exp1}-\cite{exp7}, as well
as in other photonic media, including exciton-polariton condensates \cite%
{exci2,exci3}, and in optomechanical systems \cite{om}. In these contexts,
breaking of the $\mathcal{PT}$ symmetry was observed. Emulation of the $%
\mathcal{PT}$ symmetry was also demonstrated in acoustics \cite{acoustics}
and electronic circuits \cite{electronics}, and predicted in atomic
Bose-Einstein condensates \cite{Cartarius}, magnetism \cite{magnetism}, and
chains of coupled pendula \cite{Peli}.

The $\mathcal{PT}$ symmetry, being a linear feature, is often combined with
intrinsic nonlinearity of settings in which it is realized. Most typically,
it is the Kerr nonlinearity of underlying optical media, which gives rise to
nonlinear Schr\"{o}dinger equations (NLSEs) with the cubic term and complex
potentials, subject to constraint Eq. (\ref{minus}). Such equations may
generate $\mathcal{PT}$-symmetric solitons, which were considered in many
theoretical works \cite{soliton}, \cite{Konotop}-\cite{Alexeeva} (see also
reviews \cite{review1,review2}), and experimentally demonstrated too \cite%
{exp7}. Although these works were chiefly dealing with one-dimensional (1D)
models, stable $\mathcal{PT}$-symmetric solitons were also predicted in some
2D settings \cite{Yang}, \cite{2D-1}-\cite{2D-3}. A characteristic feature
of $\mathcal{PT}$-symmetric solitons is that, although existing in
dissipative systems, they appear in continuous families, similar to their
counterparts in conservative models \cite{families}, while usual dissipative
solitons exist as isolated solutions (\textit{attractors}, if they are
stable) \cite{diss2,diss3}. The realization of the $\mathcal{PT}$ symmetry
in 2D geometry may provide essential extension of the above-mentioned
applications, such as the unidirectional transmission, enhanced absorption,
and lasing for broad optical beams.

Solitons are also vulnerable to destabilization via the $\mathcal{PT}$%
-symmetry breaking at the critical value of the gain-loss coefficient \cite%
{breaking}. Nevertheless, it was found that, in some settings, the solitons'
$\mathcal{PT}$ symmetry can be made \emph{unbreakable}, extending to
arbitrarily large values of the strength of the model's imaginary potential
\cite{unbreakable}-\cite{China}, see also a brief review of the
unbreakability concept in \cite{book}. The particular property of these
models is that self-trapping of solitons is provided not by the
self-focusing sign of the nonlinearity, but by the defocusing sign, with the
coefficient in front of the cubic term growing fast enough from the center
to periphery. In the absence of gain and loss, this scheme of stable
self-trapping was elaborated for 1D, 2D, and 3D bright solitons \cite%
{Barcelona1}-\cite{Barcelona5}. It is essential to stress that such models
are \textit{nonlinearizable}, which means that decaying tails of solitons
are determined by the full nonlinear equation. In other words, the models
have no linear spectrum, the spectrum of eigenstates being represented by
nonlinear self-trapped modes (solitons). Accordingly, the models elaborated
in Refs. \cite{unbreakable}-\cite{China} realize the $\mathcal{PT}$ symmetry
in a sense different from that defined in the usual systems---not in terms
of the linear spectrum, which does not exist in this case, but in the form
of stable families of complex-valued solitons with real propagation
constants (eigenvalues), which exist in the presence of spatially odd
imaginary potentials.

The present work introduces 2D models which maintain stable solitons,
including (nearly) unbreakable ones, in the presence of the spatially
growing self-defocusing nonlinearity and antisymmetric imaginary potentials,
$iW\left( x,y\right) $ in Eq. (\ref{U}). One model, with
\begin{equation}
W(x,y)=\gamma _{0}x\exp \left( -\beta r^{2}\right) ,~r^{2}=x^{2}+y^{2},
\label{x}
\end{equation}%
where $\gamma _{0}>0$ and $\beta \geq 0$ are constants, features the
unbreakable or nearly unbreakable 2D $\mathcal{PT}$ symmetry, represented by
several species of families of stable solitons: single- and double-peak
ones, as well as 2D solitons with embedded integer vorticity (topological
charge), $m=1,2,3$. The second model, with
\begin{equation}
W\left( x,y\right) =\gamma _{0}xy\exp \left( -\beta r^{2}\right) ,
\label{xy}
\end{equation}%
is not, strictly speaking, a $\mathcal{PT}$-symmetric one, but it is equally
relevant for the realization in optics, and it shares basic manifestations
of the $\mathcal{PT}$ symmetry, maintaining families of single- and
multi-peak solitons [featuring up to five peaks, in accordance with the
structure of $W(x,y)$] and solitary vortices, also with $m=1,2,3$. The
latter result is a contribution to the general topic of constructing models
more general than the $\mathcal{PT}$-symmetric ones with similar
properties(including the case of the partial $\mathcal{PT}$ symmetry \cite%
{partial}), which has been addressed in various settings \cite%
{2D-0,2D-02,2D-03,zezyu,families,Kominis1,Kominis2,Nixon,Kominis3}, see also
review \cite{review1}.

In both models, universal analytical forms are obtained for tails of
solitons, and full exact solutions are produced for particular species of
single-peak solitons, with $\beta =0$ in Eqs. (\ref{x}) and (\ref{xy}). In
the former case, the existence of the exact solitons at arbitrarily large
values of $\gamma _{0}$ in Eq. (\ref{x}) explicitly demonstrates the
unbreakability of the $\mathcal{PT}$ symmetry. In the latter case two
different families of exact solutions are found, which, however, exist only
for $\gamma _{0}\leq 2$ in Eq. (\ref{xy}) with $\beta =0$. In addition, an
anisotropic version of the latter model gives rise to particular exact
solutions for vortex solitons with topological charge $m=1$. Generic soliton
families with $m=0,1,2,3$, which include the exact single-peak solutions as
particular ones, are constructed in a numerical form in both models, and
their stability is investigated numerically---both through computation of
eigenvalues for small perturbations and by means of direct simulations.

% The rest of the paper is structured as follows. The models and analytical
% solutions, both universal asymptotic ones for tails of solitons and
% particular exact soliton solutions, are introduced in Section II, numerical
% results for fundamental (zero-vorticity) and vortical solitons in both
% models are reported, severally, in Sections III and IV, and the paper is
% concluded by Section V.

\section*{Results}

\subsection*{\textbf{The models and analytical solutions for solitons}}

\subsubsection*{\textbf{The underlying equations}}

The 1D NLSE for the amplitude of the electromagnetic field, $u(x,z)$, with
the local strength of the self-defocusing nonlinearity, $\Sigma (x)$,
growing from $x=0$ towards $x=\pm \infty $ faster than $|x|$ (this condition
is necessary for self-trapping imposed by the self-repulsion \cite%
{Barcelona1}), which is capable to maintain bright solitons with unbreakable$%
\mathcal{\ PT}$ symmetry, is \cite{unbreakable}%
\begin{equation}
i\frac{\partial u}{\partial z}+\frac{1}{2}\frac{\partial ^{2}u}{\partial
x^{2}}-\Sigma (x)|u|^{2}u=iW(x)u.  \label{qq}
\end{equation}%
Here $z$ and $x$ are scaled propagation coordinate and transverse
coordinate, in terms of the planar optical waveguide. In work \cite%
{unbreakable}, the analysis was presented for a steep 1D modulation profile,%
\begin{equation}
\Sigma (x)=\left( 1+\sigma x^{2}\right) \exp \left( x^{2}\right) ,
\label{Gauss}
\end{equation}%
with $\sigma \geq 0$, where coefficients equal to $1$ may be fixed to these
values by means of rescaling. The choice of this profile allows one to
obtain a particular exact solution for solitons \cite{Barcelona1}. Of
course, in a real physical medium the local strength of the nonlinearity,
defined as per Eq. (\ref{Gauss}), cannot grow to infinitely large values at $%
|x|\rightarrow \infty $. However, in reality it is sufficient that it grows
according to Eq. (\ref{Gauss}) to finite values, that correspond to $|x|$
which is essentially larger than the width of the soliton created by this
profile. The growth of $\Sigma (x)$ may be safely aborted at still larger $%
|x|$ \cite{Barcelona1}.

Further, the spatially-odd imaginary potential, which accounts for the $%
\mathcal{PT}$-symmetric gain-loss profile (cf. Eq. (\ref{U})), was
introduced in Ref. \cite{unbreakable} as%
\begin{equation}
W(x)=\gamma _{0}x\exp \left( -\beta x^{2}\right) ,  \label{gamma}
\end{equation}%
with $\gamma _{0}>0$ and $\beta \geq 0$. In the case of the spatially
uniform self-focusing cubic nonlinearity, the 1D imaginary potential in the
form given by Eq. (\ref{gamma}) was introduced in Ref. \cite{extra-China}.

Here, we aim to introduce a 2D extension of the model, as the NLSE for the
propagation of the electromagnetic field with amplitude $u\left(
x,y,z\right) $ in the bulk waveguide with transverse coordinates $\left(
x,y\right) $:

\begin{equation}
i\frac{\partial u}{\partial z}+\frac{1}{2}\left( \frac{\partial ^{2}u}{%
\partial x^{2}}+\frac{\partial ^{2}u}{\partial y^{2}}\right) -\Sigma
(r)|u|^{2}u=iW\left( x,y\right) u,  \label{NLS}
\end{equation}%
where $r\equiv \sqrt{x^{2}+y^{2}}$ is the radial coordinate, and the
nonlinearity-modulation profile is chosen similar to its 1D counterpart (\ref%
{Gauss}):
\begin{equation}
\Sigma (r)=\left( 1+\sigma r^{2}\right) \exp \left( r^{2}\right)
\label{sigma}
\end{equation}%
with $\sigma \geq 0$. Further, we consider two different versions of the 2D
imaginary potential. First, it is a $\mathcal{PT}$-symmetric one given by
Eq. (\ref{x}). The other imaginary potential, defined as per Eq. (\ref{xy}),
is not $\mathcal{PT}$-symmetric, because the $\mathcal{P}$ transformation, $%
\left( x,y\right) \rightarrow \left( -x,-y\right) $, does not reverse the
sign of $W\left( x,y\right) $, in this case. However, in terms of the
implementation in optics the gain-loss distribution corresponding to Eq. (%
\ref{xy}) is as relevant as that defined by Eq. (\ref{gamma}), and, as
mentioned above, properties of solitons in models which are akin to $%
\mathcal{PT}$-symmetric ones is a subject of considerable interest.

Stationary states with a real propagation constant, $k$, are looked for as
solutions to Eq. (\ref{NLS}) in the form of
\begin{equation}
u\left( x,y\right) =\exp \left( ikz\right) U\left( x,y\right) ,  \label{uU}
\end{equation}%
with complex function $U\left( x,y\right) $ satisfying the following
equation:%
\begin{equation}
kU=\frac{1}{2}\left( \frac{\partial ^{2}U}{\partial x^{2}}+\frac{\partial
^{2}U}{\partial y^{2}}\right) -\Sigma (r)|U|^{2}U-iW\left( x,y\right) U.
\label{UU}
\end{equation}

\subsubsection*{\textbf{Asymptotic solutions}}

As mentioned above, Eqs. (\ref{NLS}) and (\ref{UU}) are \textit{%
nonlinearizable}, i.e., they cannot be characterized by a linear spectrum.
Indeed, straightforward analysis of Eq. (\ref{UU}) demonstrates that it may
produce localized solutions (solitons), with tails decaying at $r\rightarrow
\infty $ according to an asymptotic expression which is determined by the
full nonlinear equation, rather than by its linearization. For the $\mathcal{%
PT}$-symmetric imaginary potential (\ref{x}) with $\beta =0$, it is%
\begin{equation}
U_{\mathrm{asympt}}\left( x,y\right) =\frac{1}{\sqrt{2\sigma }}\exp \left( -%
\frac{1}{2}r^{2}-i\gamma _{0}x\right) ,  \label{asympt1}
\end{equation}%
provided that $\sigma \neq 0$. In the case case of $\sigma =0$, this
asymptotic solution is replaced by%
\begin{equation}
U_{\mathrm{asympt}}\left( x,y\right) =\frac{r}{\sqrt{2}}\exp \left( -\frac{1%
}{2}r^{2}-i\gamma _{0}x\right) .  \label{asympt2}
\end{equation}
Note that asymptotic solutions given by Eqs. (\ref{asympt1}) and (\ref%
{asympt2}) exist at \emph{arbitrarily large} $\gamma _{0}$, suggesting the
\textit{unbreakability} of the $\mathcal{PT}$ symmetry in this case, as
corroborated by exact solution (\ref{exact1}) produced below.

The imaginary potential defined by Eq. (\ref{xy}) with $\beta =0$ produces
the following result:%
\begin{equation}
U_{\mathrm{asympt}}\left( x,y\right) =\sqrt{\frac{1-\left( \gamma
_{0}/2\right) ^{2}}{2\sigma }}\exp \left( -\frac{1}{2}r^{2}-\frac{1}{2}%
i\gamma _{0}xy\right) ,  \label{asympt3}
\end{equation}%
for $\sigma \neq 0$, and if $\sigma =0$, the result is%
\begin{equation}
U_{\mathrm{asympt}}\left( x,y\right) =\sqrt{\frac{1-\left( \gamma
_{0}/2\right) ^{2}}{2}}r\exp \left( -\frac{1}{2}r^{2}-\frac{1}{2}i\gamma
_{0}xy\right) .  \label{asympt4}
\end{equation}%
On the contrary to the the above asymptotic solutions, given by Eqs. (\ref%
{asympt1}) and (\ref{asympt2}), which are available for arbitrarily large $%
\gamma _{0}$, their counterparts produced by Eqs. (\ref{asympt3}) and (\ref%
{asympt4}) exist only at $\gamma _{0}<2$, i.e., if the gain-loss coefficient
is not too large.

It is relevant to stress the \textit{universal character} of all asymptotic
approximations given by Eqs. (\ref{asympt1}) - (\ref{asympt4}): they depend
solely on coefficients $\sigma $ and $\gamma _{0}$ of the underlying model,
and, unlike the commonly known asymptotic forms of solitons in usual
systems, do not depend on the propagation constant, $k$. The single
exception is presented by exact solution Eq. (\ref{exact0}) given below,
whose asymptotic form (actually coinciding with the exact soliton solution,
in that case) explicitly depends on $k$, but this happens solely for
specially chosen parameters given by Eq. (\ref{special}). In the generic
case, a dependence on $k$ appears in the next-order correction to the shape
of the asymptotic tail. In particular, the correction to the tails given by
Eqs. (\ref{asympt1}) and (\ref{asympt2}) are%
\begin{equation}
\delta U_{\mathrm{asympt}}\left( x,y\right) =-\left( k/r^{2}\right) U_{%
\mathrm{asympt}}\left( x,y\right) .  \label{delta}
\end{equation}%
Furthermore, for more complex solutions, such as multi-peak solitons and
solitary vortices, as well as for higher-order radial states of the
single-peak solitons, which are produced below in the numerical form, the
asymptotic form at large $r$ is exactly the same as given by Eqs. (\ref%
{asympt1})-(\ref{asympt4}).

\subsubsection*{\textbf{Exact solutions for single-peak solitons}}

Precisely at the above-mentioned critical value $\gamma _{0}=2$, the
asymptotic solutions (\ref{asympt3}) and (\ref{asympt4}) vanish. However, in
the special case,
\begin{equation}
\sigma =0,\gamma _{0}=2,\beta =0,  \label{special}
\end{equation}%
the vanishing asymptotic solution Eq. (\ref{asympt4}) is replaced by a
different one, which, as can be easily checked, is an \emph{exact solution}
to Eq. (\ref{UU}) (not just an asymptotic approximation valid at large $r$),%
\begin{equation}
\left( U_{\mathrm{exact}}^{\left( xy\right) }\right) _{\gamma _{0}=2}=\sqrt{%
-\left( 1+k\right) }\exp \left( -\frac{1}{2}r^{2}-\frac{1}{2}i\gamma
_{0}xy\right) .  \label{exact0}
\end{equation}%
It exists, as the continuous family, at all values of $k<-1$.

Further, Eq. (\ref{UU}) which includes the $\mathcal{PT}$-symmetric
imaginary potential Eq. (\ref{x}), with $\beta =0$, gives rise to an exact
solution at a special value $k_{0}^{(x)}$ of the propagation constant:%
\begin{equation}
U_{\mathrm{exact}}^{(x)}=\frac{1}{\sqrt{2\sigma }}\exp \left( -\frac{1}{2}%
r^{2}-i\gamma _{0}x\right) ,  \label{exact1}
\end{equation}%
\begin{equation}
~k_{0}^{(x)}=-\left( 1+\frac{\gamma _{0}^{2}}{2}+\frac{1}{2\sigma }\right) ,
\label{exact1parameters}
\end{equation}%
which exists at all values of coefficients $\gamma _{0}$ and $\sigma $,
except for $\sigma =0$. In other words, at $k=k_{0}^{(x)}$ the asymptotic
approximation Eq. (\ref{asympt1}) is tantamount to the exact solution. This
solution features the \emph{unbreakable} $\mathcal{PT}$ symmetry, as it
persists at arbitrarily large values of the gain-loss coefficient, $\gamma
_{0}$. Moreover, although Eq. (\ref{exact1}) yields the exact solution at
the single value of the propagation constant, given by Eq. (\ref%
{exact1parameters}), which is embedded in a generic family of numerically
found fundamental solitons, as demonstrated below in Figs. \ref{fig1}-\ref%
{fig3}, the entire family asymptotically shrinks to the exact solution in
the limit of large $\gamma _{0}$. Indeed, it is easy to find that, for $%
\gamma _{0}^{2}\gg 1$ and a relatively small deviation of the propagation
constant from the special value (\ref{exact1parameters}), $\left\vert \delta
k\right\vert \equiv \left\vert k-k_{0}^{(x)}\right\vert \ll \gamma _{0}^{2}$%
, the fundamental soliton is%
\begin{equation}
U_{\mathrm{approx}}^{(x)}\approx \frac{1}{\sqrt{2\sigma }}\exp \left[ -\frac{%
1}{2}\left( r^{2}+\frac{\delta k}{\gamma _{0}^{2}}x^{2}\right) -i\left(
\gamma _{0}-\frac{\delta k}{\gamma _{0}}\right) x\right] ,  \label{approx}
\end{equation}%
featuring weak anisotropy of the shape, $\left\vert U_{\mathrm{approx}%
}^{(x)}\left( x,y\right) \right\vert $.

Next, Eq. (\ref{UU}) with the imaginary potential taken as per Eq. (\ref{xy}%
) with $\beta =0$, and with $\sigma \neq 0$ in the nonlinearity-modulation
profile (\ref{sigma}), gives rise to the following exact solution, at the
respective single value of $k$:%
\begin{equation}
\left( U_{\mathrm{exact}}^{\left( xy\right) }\right) _{\gamma _{0}<2}=\sqrt{%
\frac{1-\left( \gamma _{0}/2\right) ^{2}}{2\sigma }}\exp \left( -\frac{1}{2}%
r^{2}-\frac{1}{2}i\gamma _{0}xy\right) ,  \label{exact2}
\end{equation}%
\begin{equation}
k_{0}^{(xy)}=-\left[ 1+\frac{1}{2\sigma }\left( 1-\left( \frac{\gamma _{0}}{2%
}\right) ^{2}\right) \right] .  \label{exact2parameters}
\end{equation}%
In this case too, the asymptotic approximation Eq. (\ref{asympt3}) becomes
identical to the exact solution at $k=k_{0}^{(xy)}$, both existing at $%
\gamma _{0}<2$, on the contrary to exact solution (\ref{exact1}), which
exists at all values of $\gamma _{0}$.

Thus, the models considered here do not have the linear spectrum. Instead of
it, they are characterized by spectra (families) of self-trapped nonlinear
solutions (solitons). The radical change of the concept of the system's
spectrum implies a respective change in the concept of the $\mathcal{PT}$
symmetry, which now applies not to the set of eigenvalues of the linearized
system, but directly to the existence of families of nonlinear states.
Lastly, it is worthy to note that all the asymptotic and exact solutions
produced above, including the first correction (\ref{delta}) to the
asymptotic tails, feature isotropic shapes of $\left\vert U(x,y\right\vert $%
, although the imaginary potentials Eqs. (\ref{x}) and (\ref{xy}) are
obviously anisotropic.

\subsubsection*{\textbf{Exact solutions for elliptic vortices in an
anisotropic model}}

In addition to 2D fundamental solitons, similar to the exact ones presented
here, we also address below, by means of numerical methods, solitons with
embedded vorticities, $m=1,2,3...$ . A challenging issue is to seek for
exact solutions for vortex solitons. Such solutions can be found in the case
of imaginary potential Eq. (\ref{xy}) with $\beta =0$, but for a more
general \textit{anisotropic} version of the nonlinearity-modulation profile
in Eq. (\ref{NLS}) with $\sigma =0$, namely,%
\begin{equation}
\Sigma \left( x,y\right) =\exp \left( x^{2}+gy^{2}\right) ,  \label{g}
\end{equation}%
where positive $g\neq 1$ accounts for the ellipticity of the modulation
profile. Then, an exact solution for elliptically deformed vortex solitons
with $m=1$ is given by the following ansatz [cf. Eq. (\ref{exact2})]:
\begin{equation}
U\left( x,y\right) =U_{0}\left( x+iby\right) \exp \left( -\frac{1}{2}\left(
x^{2}+gy^{2}\right) -iaxy\right) ,  \label{vortex}
\end{equation}%
where real $b\neq 1$ accounts for the ellipticity of the soliton's phase
field, and $a$ is another real constant. The substitution of this ansatz and
expressions Eq. (\ref{g}) and Eq. (\ref{xy}) (with $\beta =0$) in the
accordingly modified equation (\ref{UU}) leads to the following relations
between parameters of the ansatz:%
\begin{gather}
\left( 1+g\right) a=-\gamma _{0},  \notag \\
(g-1)b-\left( 1+b^{2}\right) a=0,  \label{conditions} \\
b^{2}\left( 1-a^{2}\right) +a^{2}=g^{2},  \notag
\end{gather}%
supplemented by expressions for the propagation constant and soliton's
amplitude:%
\begin{equation}
k=-\left( 3/2+g/2+ab\right) ,~U_{0}^{2}=\left( 1-a^{2}\right) /2.  \label{kU}
\end{equation}%
The system of three equations (\ref{conditions}) for two free parameters $a$
and $b$ demonstrates that the exact vortex solution is a nongeneric one, as
it may exist only if an additional constraint, which can be derived by
eliminating $a$ and $b$ in Eq. (\ref{kU}), is imposed on parameters $g$ and $%
\gamma _{0}$:%
\begin{equation}
\left( g^{2}-1\right) ^{2}\left[ g^{2}\left( g+1\right) ^{2}-\gamma _{0}^{2}%
\right] \left[ \left( g+1\right) ^{2}-\gamma _{0}^{2}\right] =\gamma _{0}^{2}%
\left[ \left( g^{2}+1\right) \left( g+1\right) ^{2}-2\gamma _{0}^{2}\right]
^{2}.
\end{equation}

In the isotropic model, with $g=1$, Eq. (\ref{conditions}) has no nontrivial
solutions. However, they can be found for $g\neq 1$. A particular example is%
\begin{gather}
b=1/\sqrt{2}\approx 0.707\,1,a=-\left( 3-\sqrt{5}\right) /\left( 4\sqrt{2}%
\right) \approx -0.1351, \\
U_{0}=\sqrt{3\left( 3+\sqrt{5}\right) }/\left( 4\sqrt{2}\right) \approx
0.7006,
\end{gather}%
which is a valid solution at $g=\left( 3\sqrt{5}-1\right) /8\approx
\allowbreak 0.713\,5\ $and $\gamma _{0}=\left( 3+\sqrt{5}\right) /\left( 16%
\sqrt{2}\right) \approx \allowbreak 0.231\,4$. This value of $g$ corresponds
to eccentricity $e\equiv \sqrt{1-g}=\sqrt{(9-3\sqrt{5})/8}\approx
\allowbreak 0.535\,2$ of the elliptic profile in Eq. (\ref{g}).

Numerical results are reported below for the isotropic model, while the
anisotropic one should be a subject for separate consideration.

\subsection*{\textbf{Numerical results for zero-vorticity solitons}}

\subsubsection*{The\textbf{\ }$\mathcal{PT}$\textbf{-symmetric imaginary
potential (\protect\ref{x}): single- and double-peak solitons}}

The isolated exact solution of the model with the $\mathcal{PT}$-symmetric%
\textbf{\ }gain-loss distribution, given by Eqs. (\ref{exact1}) and (\ref%
{exact1parameters}), can be embedded in a continuous family of solitons,
produced by a numerical solution of Eq. (\ref{UU}), with $\Sigma (r)$ and $%
\gamma \left( x\right) $ taken as per Eqs. (\ref{sigma}) and (\ref{x}). The
appropriate numerical algorithm is the Newton conjugate gradient method \cite%
{Yang-book}, which is briefly outlined in section Method below. The
stability of the stationary states was identified by numerical computation
of eigenvalues of small perturbations, using linearized equations (\ref%
{eigen}) for perturbations around the stationary solitons. Finally, the
stability predictions, produced by the eigenvalues, were verified by
simulations of the perturbed evolution of the solitons (some technical
details are reported elsewhere \cite{book}).

It is relevant to stress that the convergence of the algorithm which
produces stationary states depends on appropriate choice of the initial
guess. While stationary modes were not found in \textquotedblleft holes"
appearing in stability charts which are displayed below in Figs. \ref{fig2}, %
\ref{fig3}, \ref{fig7}, \ref{fig10}-\ref{fig12} and \ref{fig16}, \ref{fig17}%
, it is plausible that stationary solutions exist in the holes too, being,
however, especially sensitive to the choice of the input. On the other hand,
the intricate alternation of stability and instability spots, which is also
observed in the charts, is a true peculiarity of the present model.
Moreover, genuine structure of the stability charts may be fractal, but
analysis of this possibility is beyond the scope of the present work.

\begin{figure}[tbp]
\centering\includegraphics[width=0.75\textwidth]{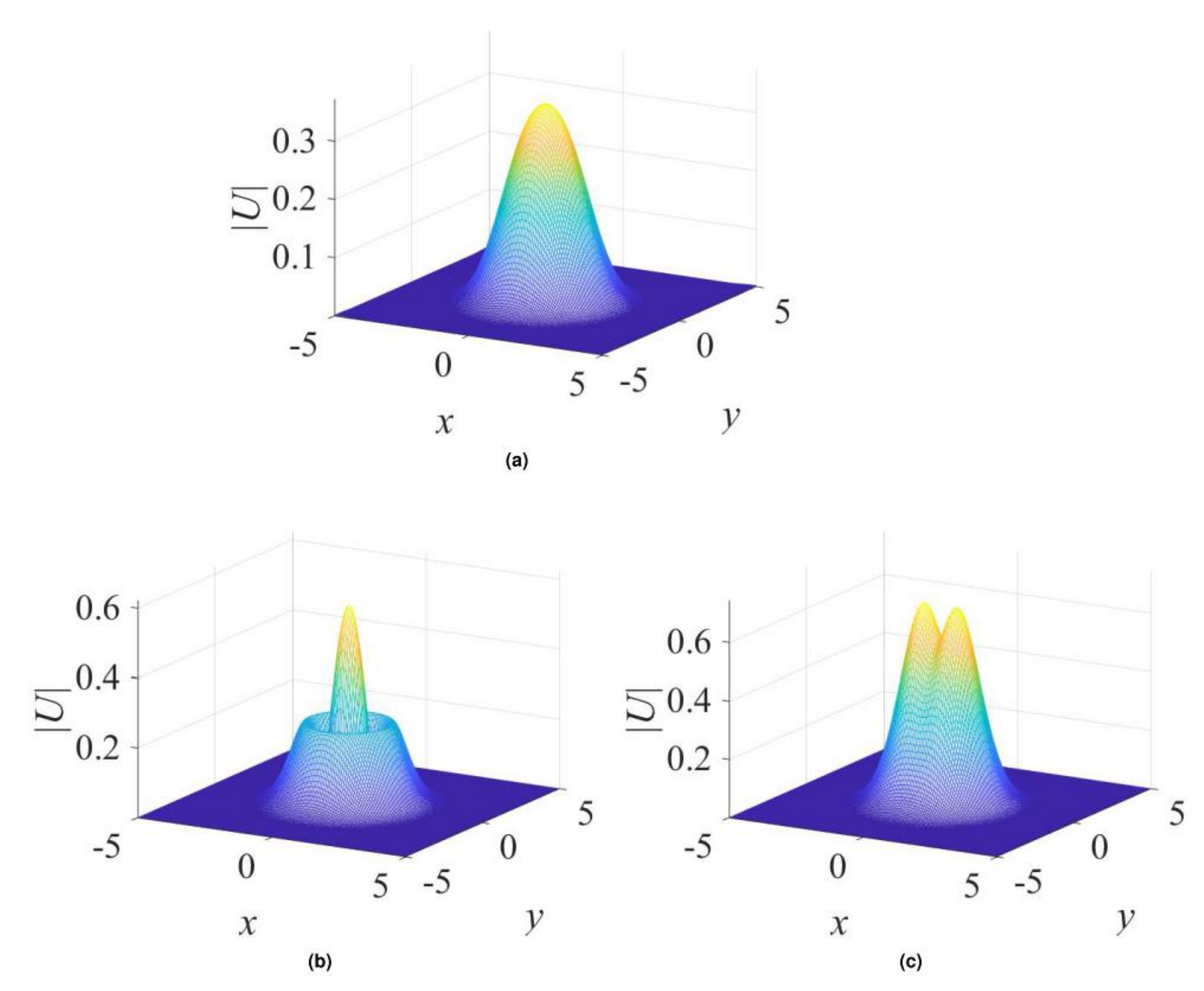}
\caption{Typical examples of stable solitons produced by the model with the $%
\mathcal{PT}$-symmetric imaginary potential defined by Eq. (\protect\ref{x}%
). (a) A fundamental single-peak soliton for $\protect\gamma _{0}=1.2$ in
Eq. (\protect\ref{x}) and propagation constant $k=-3.2$ in Eq. (\protect\ref%
{uU}). (b) A higher-order radial state of the single-peak soliton for $%
\protect\gamma _{0}=0.2$ and $k=-4$. (c) A double-peak soliton for $\protect%
\gamma _{0}=1.4$ and $k=-4$. In all the cases, $\protect\sigma =1$ and $%
\protect\beta =0$ are fixed in Eqs. (\protect\ref{sigma}) and (\protect\ref%
{x}).}
\label{fig1}
\end{figure}

Generic examples of numerically found \emph{stable} solitons with single-
and double-peak shapes are displayed in Fig. \ref{fig1}. Note that the
double-peak modes have their two maxima separated in the direction of $x$,
in accordance with the anisotropic shape of the imaginary potential in Eq. (%
\ref{x}). As concerns single-peak modes, two different varieties of stable
ones were found: fundamental solitons, with the shape similar to that of the
exact solution given by Eqs. (\ref{exact1}) and (\ref{exact1parameters})
[see Fig. \ref{fig1}(a)], and higher-order states with a radial ring
surrounding the central peak, see Fig. \ref{fig1}(b). It is worthy to note
that, unlike many other models, where higher-order radial states are
unstable \cite{Atai1}-\cite{Atai7}, they are stable in the present case.
Note also that shapes of both species of the single-peak solitons,
fundamental and higher-order ones, seem isotropic in terms of $\left\vert
U\left( x,y\right) \right\vert $, similar to exact solution (\ref{exact1}).
The isotropy is obviously broken by double-peak modes, see Fig. \ref{fig1}%
(c).

Results of the stability analysis, based on the computation of perturbation
eigenvalues, are summarized in the stability map in the plane of $\left(
k,\gamma _{0}\right) $ [the soliton's propagation constant and strength of
the gain-loss term in Eq. (\ref{x})], for $\beta =0$ and $\beta =0.2$ in
Figs. \ref{fig2} and \ref{fig3}, respectively. Several noteworthy features
are revealed by these plots. First, it is worthy to note significant
stability areas for both the double-peak and higher-order single-peak $%
\mathcal{PT}$-symmetric solitons in Figs. \ref{fig2} and \ref{fig3}.
Further, bistability is observed at many points, in the form of coexisting
stable fundamental and double-peak solitons, or fundamental and higher-order
single-peak ones. As concerns the possibility of maintaining the unbreakable
$\mathcal{PT}$ symmetry, Fig. \ref{fig2} demonstrates shrinkage of the
existence and stability regions of the modes with the increase of $\gamma
_{0}$ at $\beta =0$ to the exact soliton solution given by Eqs. (\ref{exact1}%
) and (\ref{exact1parameters}), in agreement with the trend represented by
approximate solution (\ref{approx}). Eventually, the exact solution loses
its stability at $\gamma _{0}\geq 2$. On the other hand, the introduction of
a relatively weak confinement of the gain-loss term, with $\beta =0.2$ in
Eq. (\ref{x}), demonstrates that the $\mathcal{PT}$ symmetry remains
unbreakable in \ref{fig3}, where both the existence and stability regions
extend in the direction of large values of $-k$ and $\gamma _{0}$, without
featuring any boundary.

As concerns unstable solitons, they typically blow up in the course of the
evolution, see an example below in Fig. \ref{fig18}. Although it shows the
blowup of a vortex soliton, the instability development of zero-vorticity
ones is quite similar.

\begin{figure}[tbp]
\centering\includegraphics[width=0.6\textwidth]{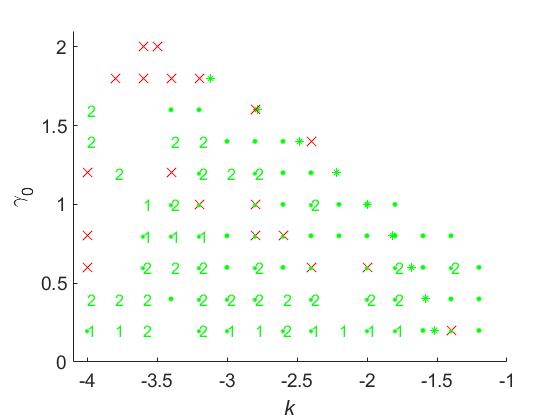}
\caption{The stability map for the $\mathcal{PT}$-symmetric solitons
maintained by imaginary potential (\protect\ref{x}), in the case of $\protect%
\sigma =1$ and $\protect\beta =0$ in Eqs. (\protect\ref{sigma}) and (\protect
\ref{x}). Stable fundamental single-peak solitons are marked by green dots.
All unstable solitons are marked by red crosses, irrespective of their
structure. Exact soliton solutions, given by Eqs. (\protect\ref{exact1}) and
(\protect\ref{exact1parameters}), are indicated by green stars (except for
one at $\protect\gamma _{0}=2$, which is designated by the red cross, as the
exact solutions are unstable at $\protect\gamma _{0}\geq 2$). Green numbers $%
\geq 2$ in this figure and below denote stable solitons with the same number
of peaks. Further, green numbers $1$ label stable single-peak solitons with
the higher-order radial structure, as in Fig. \protect\ref{fig1}(b). Green
numbers $1$ or $2$, placed close to green dots, imply bistability, i.e.,
coexistence of stable fundamental single-peak solitons and stable
higher-order or double-peak ones. Red crosses placed on top of green dots
imply coexistence of fundamental single-peak solitons with some unstable
mode. Soliton solutions were not found in white areas.}
\label{fig2}
\end{figure}

The stability charts, drawn in Figs. \ref{fig2} and \ref{fig3} for $\sigma
=1 $ in Eq. (\ref{sigma}), are quite similar to their counterparts produced
at other values of $\sigma $, including $\sigma =0$, when the exact solution
given by Eqs. (\ref{exact1}) and (\ref{exact1parameters}) does not exist,
while the asymptotic form of the solitons' tails is given by Eq. (\ref%
{asympt2}).

\begin{figure}[tbp]
\centering\includegraphics[width=0.6\textwidth]{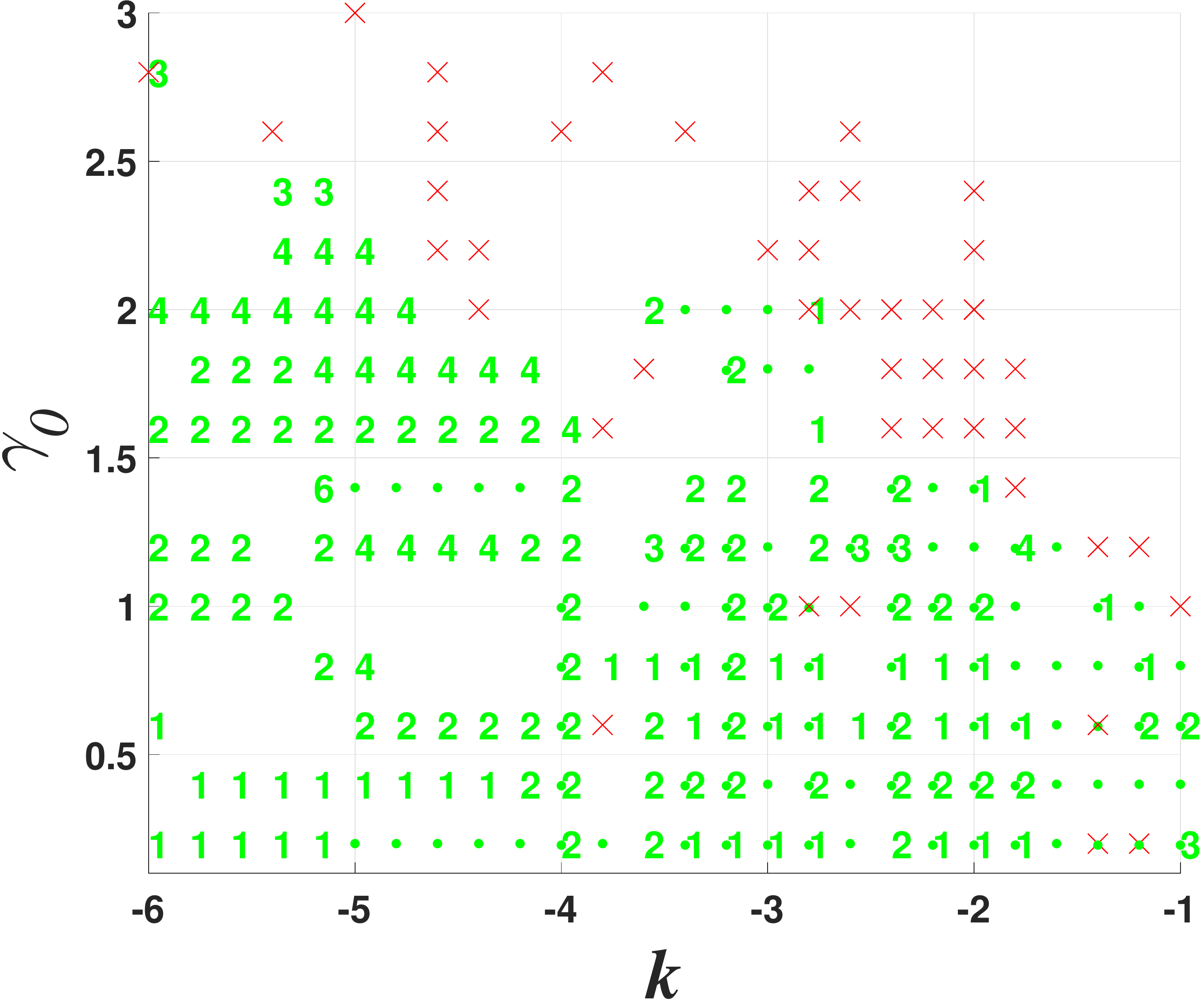}
\caption{The same as in Fig. \protect\ref{fig2}, but for $\protect\beta =0.2$
in Eq. ( \protect\ref{x}), i.e., with the gain-loss term subject to weak
spatial confinement. In this case, there are no exact solitons solutions,
while the asymptotic solution for the tails is given by Eq. (\protect\ref%
{asympt1}) with $\protect\gamma _{0}=0$ (the confinement eliminates $\protect%
\gamma _{0}$ from the asymptotic solution).}
\label{fig3}
\end{figure}

\subsubsection*{\textbf{The\ imaginary potential (\protect\ref{xy}): single-
and multi-peak solitons}}

A drastic difference revealed by the stability analysis of the model based
on Eqs. (\ref{NLS}), (\ref{sigma}) and (\ref{xy}) is that the respective
exact solutions, given by Eq. (\ref{exact0}) for the special case (\ref%
{special}), and by Eqs.\ (\ref{exact2}) and (\ref{exact2parameters}) for $%
\sigma >0$, $\beta =0$ and arbitrary $\gamma _{0}$, are completely unstable,
on the contrary to the stability of the exact solutions in the case of the $%
\mathcal{PT}$-symmetric imaginary potential Eq. (\ref{x}) (at $\gamma _{0}<2$%
). Furthermore, all numerical solutions found in the full 2D model with $%
\beta =0$ in Eq. (\ref{xy}) are unstable too. The stabilization in this
model is provided by $\beta >0$, i.e., by imposing the spatial confinement
on the gain-loss term in Eq. (\ref{xy}). For fixed $\sigma $, there is a
minimum value $\beta _{\min }$ of $\beta $ which secures the stabilization.
For instance, we have concluded that the solitons may be stable in the model
with $\sigma =1$ in Eq. (\ref{sigma}) at $\beta \geq \beta _{\min }\approx
0.2$ in Eq. (\ref{xy}), still being completely unstable, e.g., at $\beta
=0.1 $.

\begin{figure}[tbp]
\centering\includegraphics[width=0.75\textwidth]{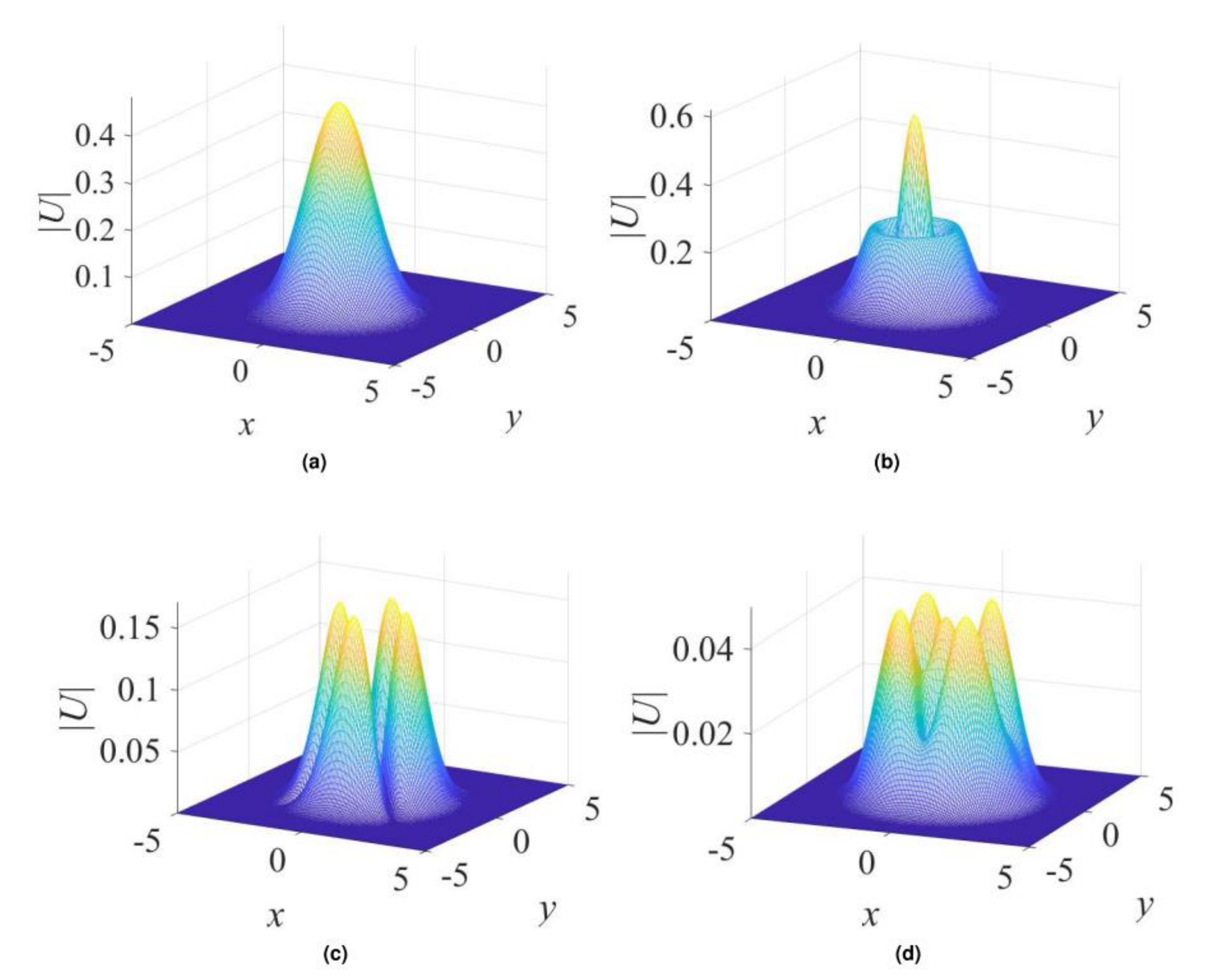}
\caption{Examples of stable single- and multi-peak $\mathcal{PT}$-symmetric
solitons, found in the model based on Eqs. (\protect\ref{sigma}) and (%
\protect\ref{xy}), with $\protect\sigma =1$ and (a) $\protect\beta =0.5$, $%
\protect\gamma _{0}=1$, $k=-1$; (b) $\protect\beta =0.5$, $\protect\gamma %
_{0}=0.2$, $k=-4$; (c) $\protect\beta =0.2$, $\protect\gamma _{0}=1.4$, $%
k=-2.8$; (d) $\protect\beta =0.5$, $\protect\gamma _{0}=0.4$, $k=-1.8$. }
\label{fig4}
\end{figure}

As mentioned above, the steep growth of $\Sigma \left( r\right) $ in Eq. (%
\ref{sigma}) cannot extend to infinity, it being sufficient to maintain the
adopted profile of $\Sigma (r)$ on a scale which is essentially larger than
a characteristic size of solitons supported by this profile. The same
pertains to the linear growth of the imaginary potential at large $|x|$ in
Eq. (\ref{x}): in reality, it should not continue at distances much larger
than the size of the stable solitons considered in the previous section.
However, the presence of $\beta _{\min }$ implies that the corresponding
\textquotedblleft tacit" confinement of $\gamma \left( x,y\right) $ in Eq. (%
\ref{xy}) is not sufficient to produce stable 2D solitons. At $\beta >\beta
_{\min }$, the numerical solution generates stable fundamental single-peak
solitons and their higher-order radial counterparts with isotropic shapes of
$\left\vert U\left( x,y\right) \right\vert $, as shown in Fig. \ref{fig4}%
(a,b). Further, stable multi-peak solitons are found too. Due to the 2D
structure of the imaginary potential (\ref{xy}), they feature a four- or
five-peak structure, built along both the $x$ and $y$ axes, as shown in Fig. %
\ref{fig4}(c,d), instead of the uniaxial double-peak modes supported by the
quasi-1D imaginary potential (\ref{x}), cf. \ref{fig1}(c).

\begin{figure}[tbp]
\centering\includegraphics[width=0.6\textwidth]{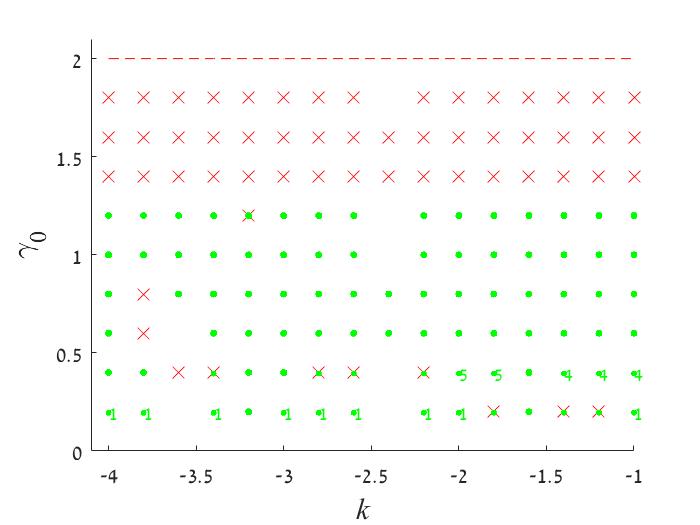}
\caption{The stability chart, defined as in Figs. \protect\ref{fig2} and
\protect\ref{fig3}, but for the model including imaginary potential (\protect
\ref{xy}), with $\protect\sigma =1$ and $\protect\beta =0.5$ in Eqs. (%
\protect\ref{sigma}) and (\protect\ref{xy}). As indicated by the upper
dashed red curve, no solitons were found at $\protect\gamma _{0}\geq 2$,
where the exact solution given by Eqs. (\protect\ref{exact2}) does not exist
either.}
\label{fig5}
\end{figure}

A typical stability chart for the 2D solitons generated by the model with $%
\beta >\beta _{\min }$ is displayed in Fig. \ref{fig5}. It features
bistability between the fundamental single-peak solitons and the
higher-order ones, or four- and five-peak complexes, in a relatively small
region of the $\left( k,\gamma _{0}\right) $ plane, at sufficiently small
values of $\gamma _{0}$. Figure \ref{fig5} clearly shows that no solitons
were found at $\gamma _{0}\geq 2$, this restriction coinciding with that for
the exact solution given by Eqs. (\ref{exact2}) and (\ref{exact2parameters}%
). Thus, unlike the $\mathcal{PT}$-symmetric imaginary potential (\ref{x}),
the model based on potential (\ref{xy}) does not produce unbreakable soliton
families.

\subsection{\textbf{Vortex solitons}}

Soliton solutions of Eq. (\ref{UU}) with embedded vorticity were found
numerically by means of the above-mentioned Newton conjugate gradient
method, initialized by the ansatz with integer vorticity $m\geq 1$ added to
the previously found 2D stationary solutions of Eq. (\ref{UU}):

\begin{equation}
U\left( x,y\right) \rightarrow U\left( x,y\right) r^{m}\exp (im\theta
)\equiv U\left( x,y\right) \left( x+iy\right) ^{m},  \label{AngularPush}
\end{equation}%
where $\left( r,\theta \right) $ are the polar coordinates. The stability of
resulting vortex solitons was again analyzed through the computation of
eigenvalues for modes of small perturbations around the vortex states, see
Eqs. (\ref{eigen}), and then verified by direct simulations.

\subsubsection*{\textbf{Vortex solitons in the case of the }$\mathcal{PT}$%
\textbf{-symmetric imaginary potential}}

\begin{figure}[tbp]
\centering\includegraphics[width=0.75\textwidth]{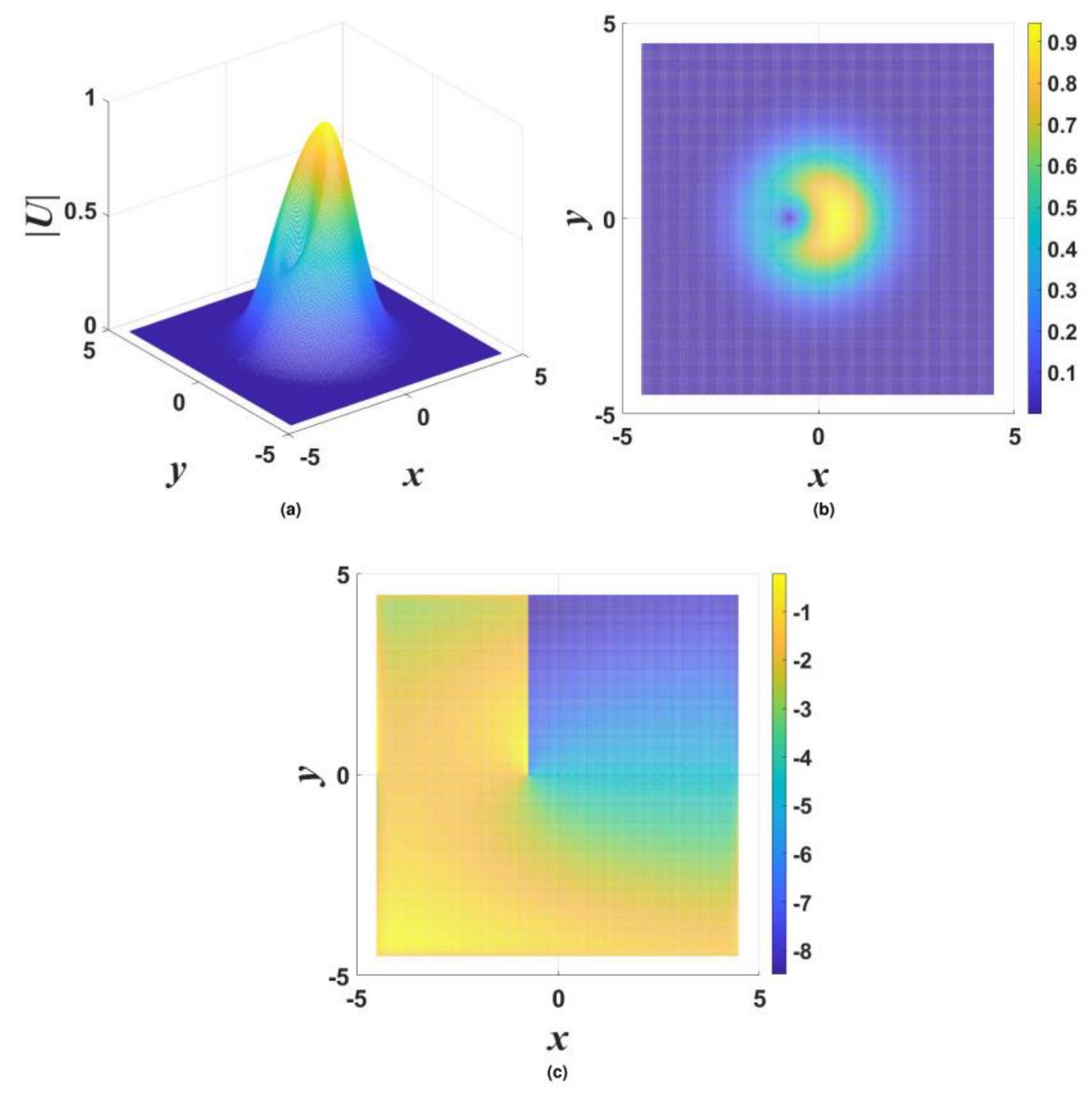}
\caption{Three-dimensional (a) and top-view (b) shapes of $\left\vert
U\left( x,y\right) \right\vert $ for a typical stable vortex soliton with $%
m=1$, supported by the $\mathcal{PT}$-symmetric imaginary potential (\protect
\ref{x}) with $\protect\gamma _{0}=0.6$, $\protect\beta =0$, and $\protect%
\sigma =0$ in Eq. (\protect\ref{sigma}), the propagation constant being $%
k=-3 $. Panel (c) displays the phase structure of the vortex.}
\label{fig6}
\end{figure}

\begin{figure}[tbp]
\centering
\centering\includegraphics[width=0.75\textwidth]{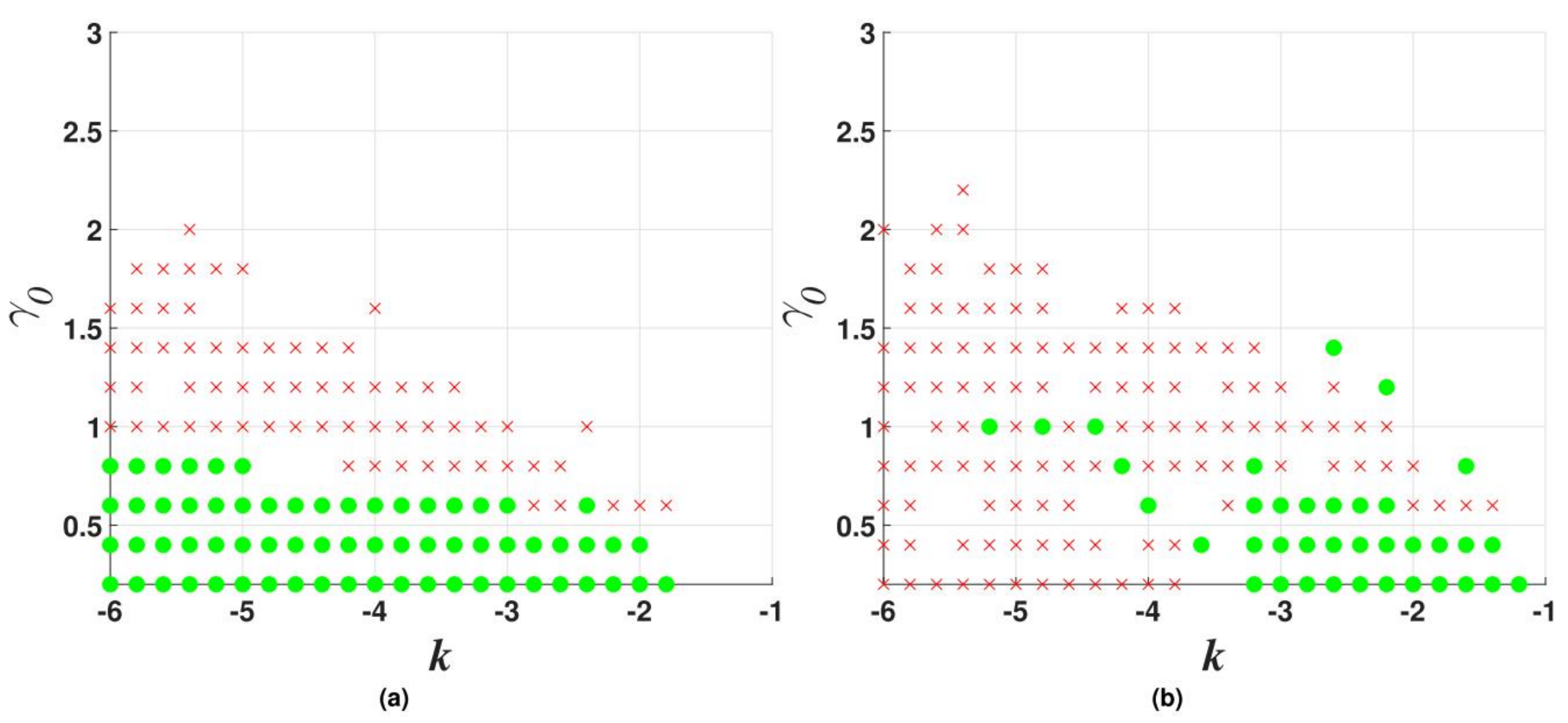}
\caption{Stability charts for vortex solitons with topological charge $m=1$
in the model including the $\mathcal{PT}$-symmetric imaginary potential (%
\protect\ref{x}) with $\protect\beta =0$, and $\protect\sigma =0$ or $1$ in
Eq. (\protect\ref{sigma}), in panels (a) and (b) panels, respectively. Green
circles and red crosses denote stable and unstable vortex solitons,
respectively. The same notation is used below in other stability charts for
vortex solitons.}
\label{fig7}
\end{figure}

In the framework of the model with imaginary potential (\ref{x}), stable
vortex solitons were found in the case of $\beta =0$ (no gain-loss
confinement) with $m=1$, while vortices with $m\geq 2$ do not exist or are
unstable. An example of stable vortices is shown in Fig. \ref{fig6}, and the
respective stability charts for different values of $\sigma $ in Eq. (\ref%
{sigma}) are presented in Fig. \ref{fig7}. The strongly anisotropic shape of
the vortex is a consequence of the anisotropy of the underlying imaginary
potential (\ref{x}).

\begin{figure}[tbp]
\centering\includegraphics[width=0.75\textwidth]{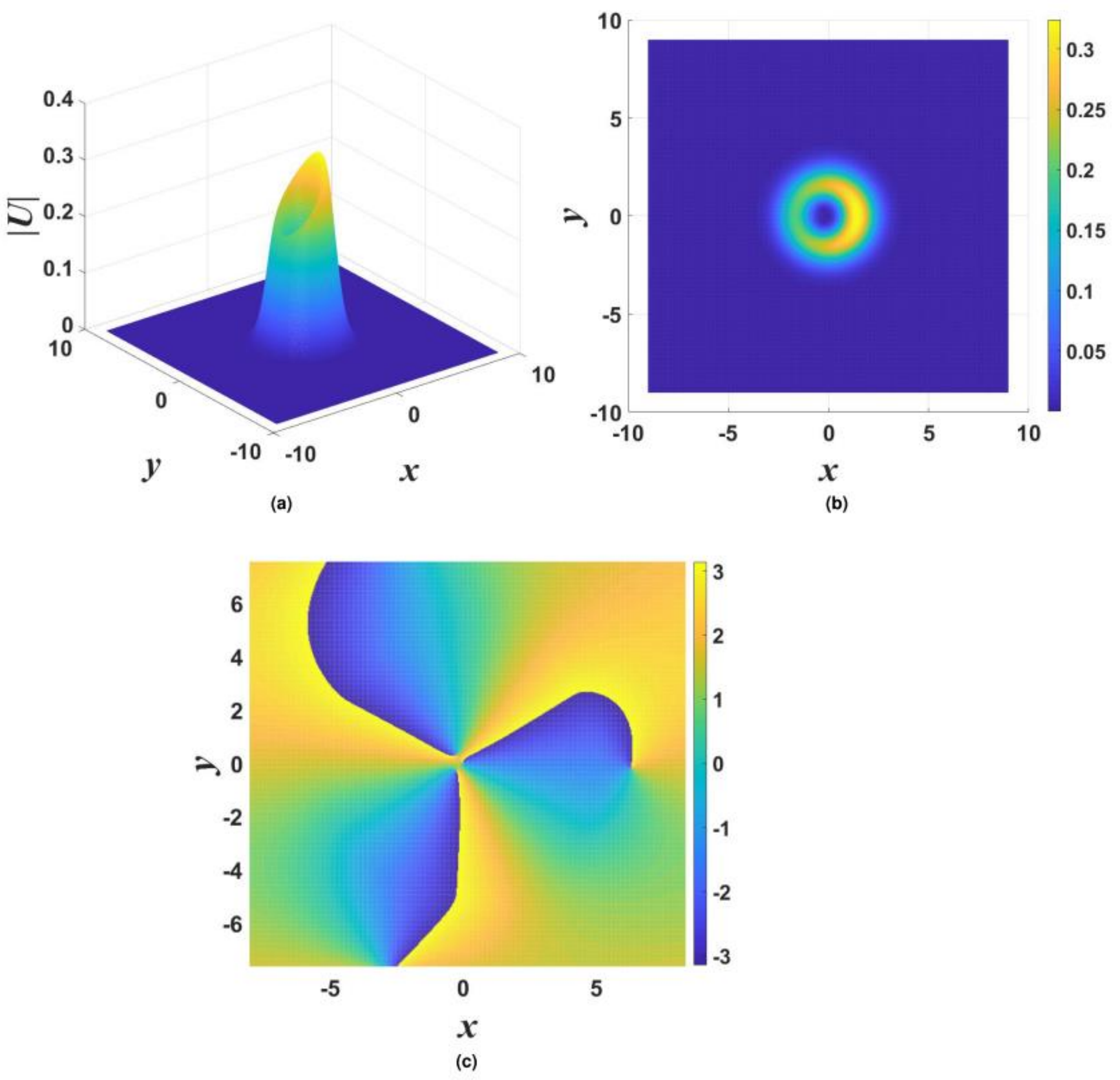}
\caption{The same as in Fig. \protect\ref{fig6}, but for stable vortex
soliton with $m=3$ and parameters $\protect\gamma _{0}=0.8$, $\protect\beta %
=0.5$, $\protect\sigma =0$, $k=-4$.}
\label{fig8}
\end{figure}

The introduction of the confinement of the gain and loss in Eq. (\ref{x})
(in particular, setting $\beta =0.5$) makes it possible to construct stable
vortex solitons with higher vorticities, corresponding to $m>1$ in Eq. (\ref%
{AngularPush}). An example of a stable vortex with $m=3$ is shown in Fig. %
\ref{fig8}.

\begin{figure}[tbp]
\centering\includegraphics[width=0.75\textwidth]{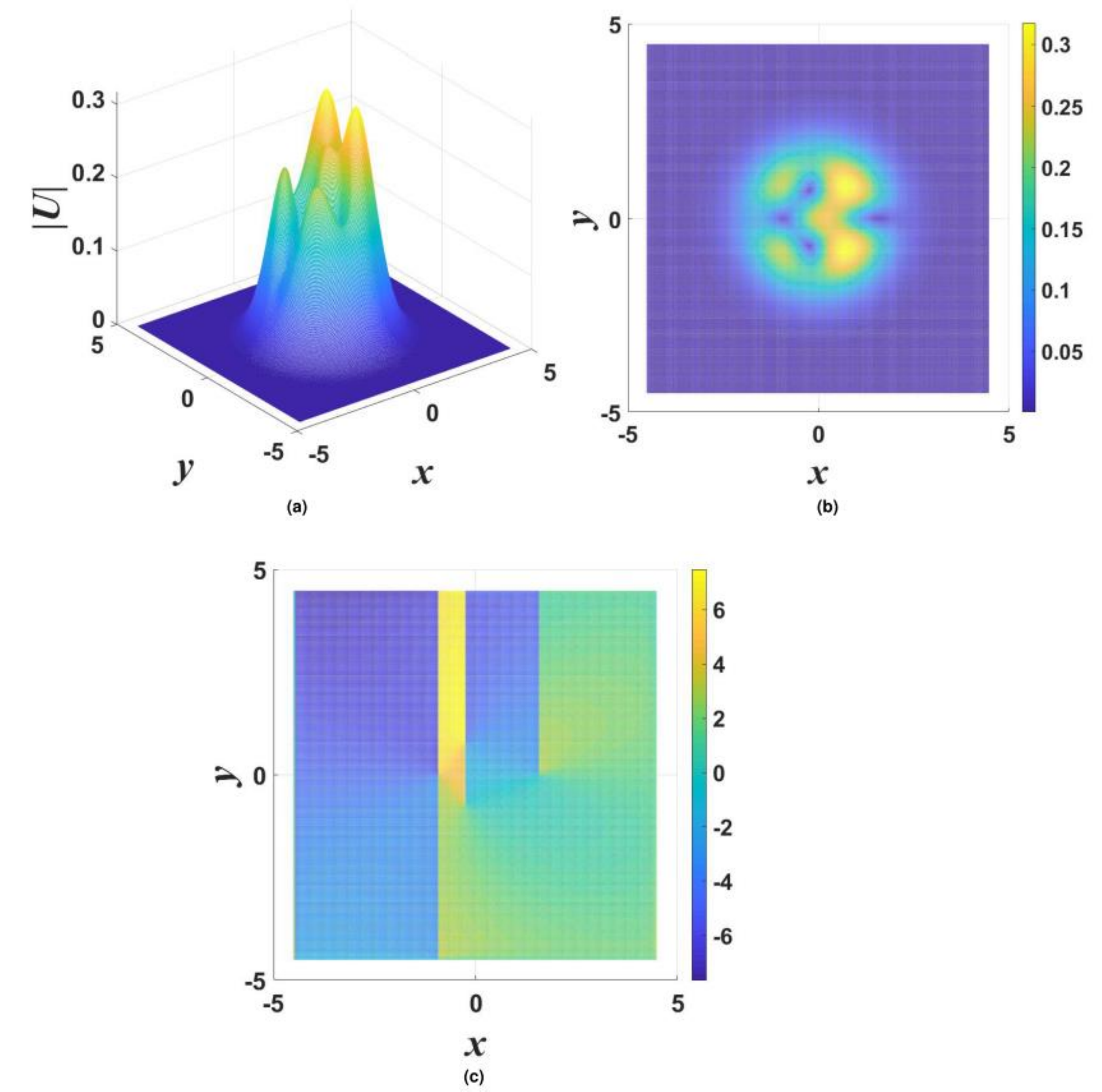}
\caption{The same as in Fig. \protect\ref{fig6}, but for a case when the
stable vortex soliton with $m=1$ and a complex shape is created, the
parameters in Eqs. (\protect\ref{x}) and (\protect\ref{sigma}) being $%
\protect\gamma _{0}=0.4$, $\protect\beta =0$, and $\protect\sigma =1$. The
propagation constant is $k=-3.6$.}
\label{fig9}
\end{figure}

In most cases, stable vortices generated by input (\ref{AngularPush}) from
double-peak stationary solutions have the same shape as those originating
from their single-peak counterparts. However, in few cases the application
of the lowest vorticity, with $m=1$ in Eq. (\ref{AngularPush}), to the
double-peak input leads to the creation of stable vortex solitons with a
complex shape, see an example in Fig. \ref{fig9}.

\begin{figure}[tbp]
\centering\includegraphics[width=0.75\textwidth]{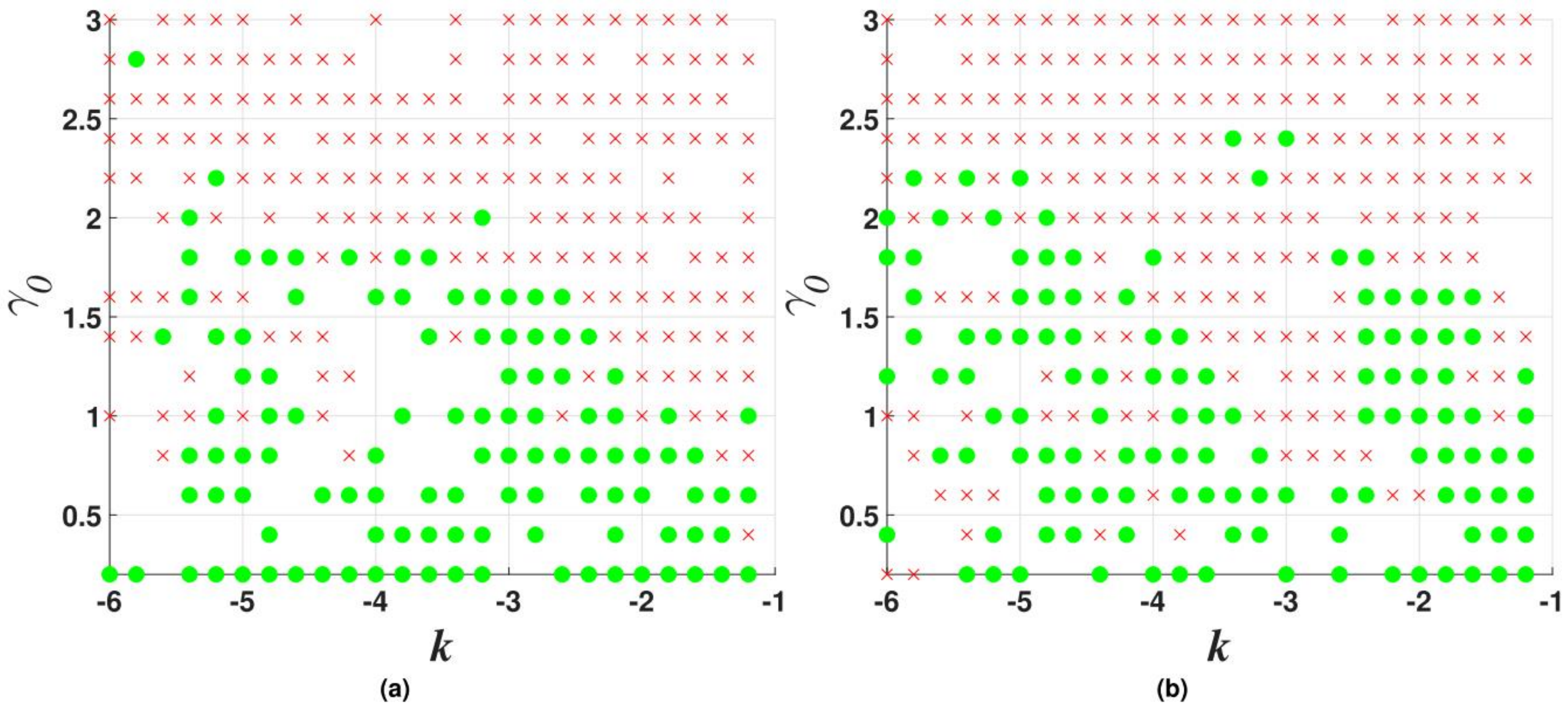}
\caption{Stability charts for solitons with vorticity $m=1$ in the case of
the $\mathcal{PT}$-symmetric imaginary potential (\protect\ref{x}) with $%
\protect\beta =0.5$, and $\protect\sigma =0$ or $1$ in Eq. (\protect\ref%
{sigma}), in panels (a) and (b), respectively.}
\label{fig10}
\end{figure}

\begin{figure}[tbp]
\centering
\centering\includegraphics[width=0.75\textwidth]{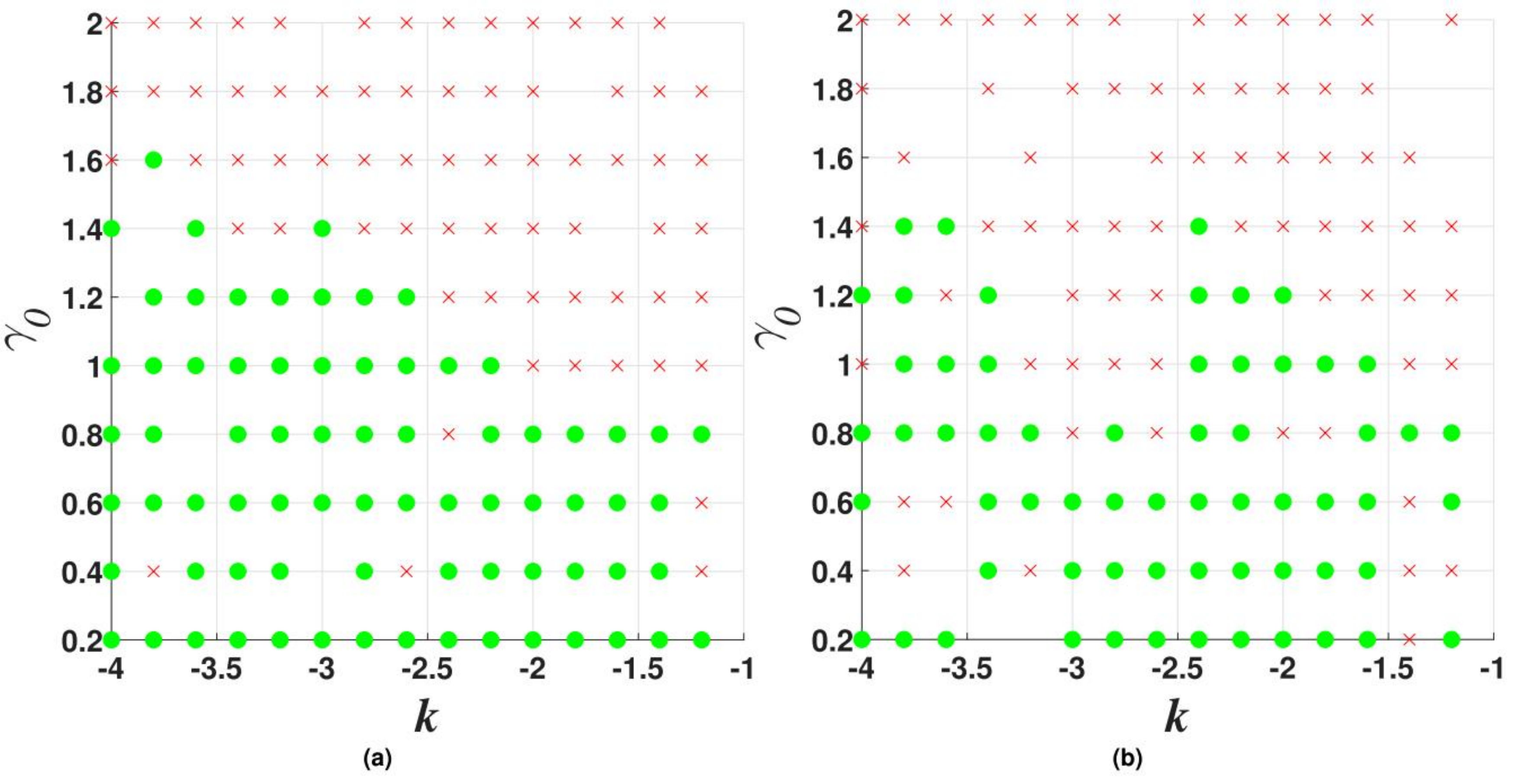}
\caption{The same as in Fig.(\protect\ref{fig10}) (stability charts) but for
vortex solitons with $m=2$.}
\label{fig11}
\end{figure}

\begin{figure}[tbp]
\centering
\centering\includegraphics[width=0.75\textwidth]{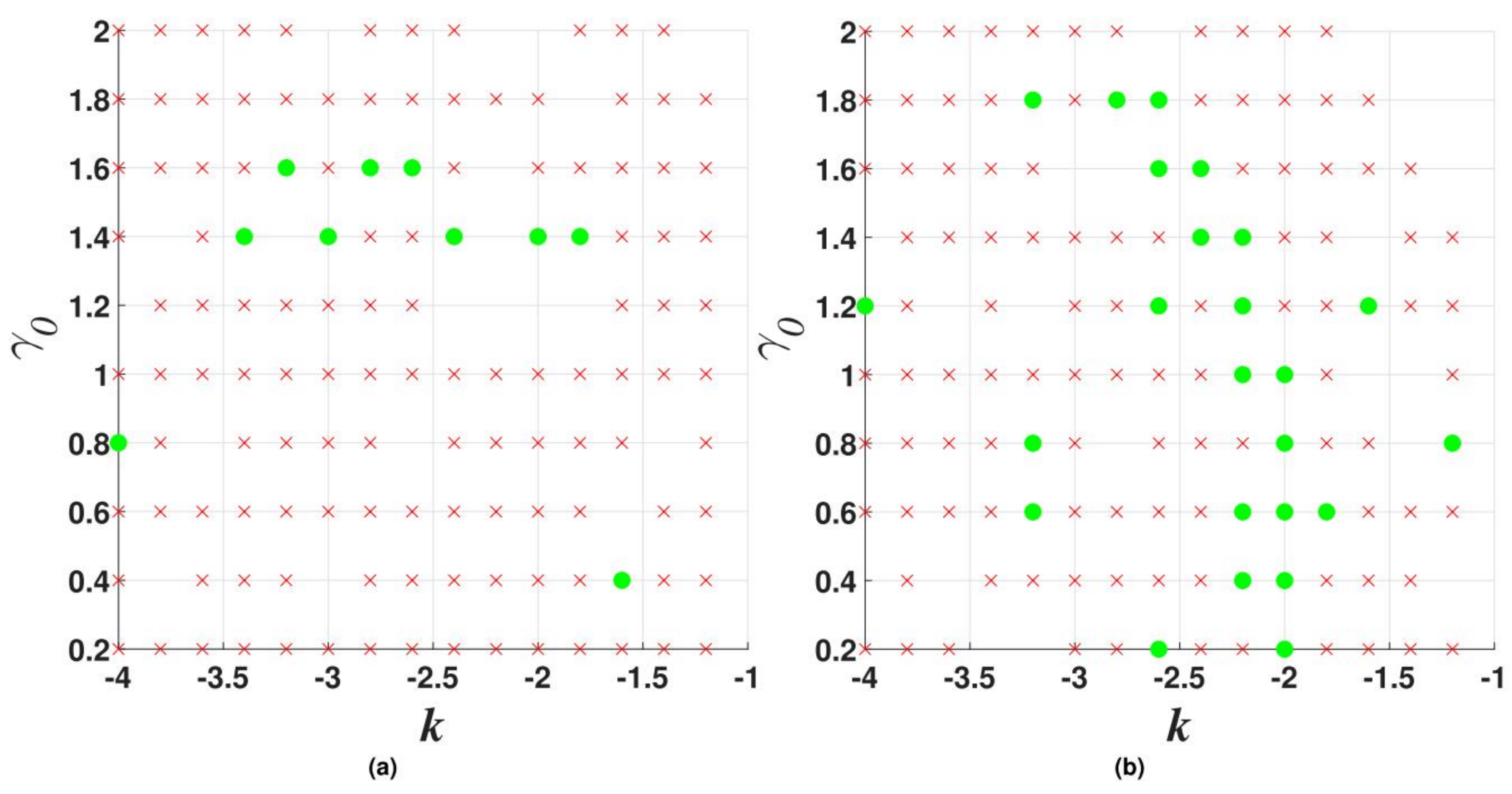}
\caption{The same as in Fig. (\protect\ref{fig10}), but for vorticity $m=3$.}
\label{fig12}
\end{figure}

Stability charts for the vortex solitons with $m=1,2$, and $3$, supported by
the $\mathcal{PT}$-symmetric imaginary potential which is subject to the
spatial confinement, with $\beta =0.5$ in Eq. (\ref{x}), are shown in Figs. %
\ref{fig10} - \ref{fig12}. While the stability area shrinks with the
increase of $m$, a few stable isolated modes were found even for $m=4$ (not
shown here). The comparison of Figs. \ref{fig10} and \ref{fig7} shows that
the introduction of the spatial confinement of the gain-loss profile helps
to expand the stability area for $m=1$ towards larger values of $\gamma _{0}$%
, thus upholding the trend to observe the unbreakable $\mathcal{PT}$
symmetry in this 2D model. In direct simulations, the evolution of unstable
vortex modes leads towards the blowup, via their fusion into a single peak,
similar to what is displayed below in Fig. \ref{fig18}.

\subsubsection*{\textbf{Vortex solitons in the model with imaginary
potential (\protect\ref{xy})}}

\begin{figure}[tbp]
\centering\includegraphics[width=0.75\textwidth]{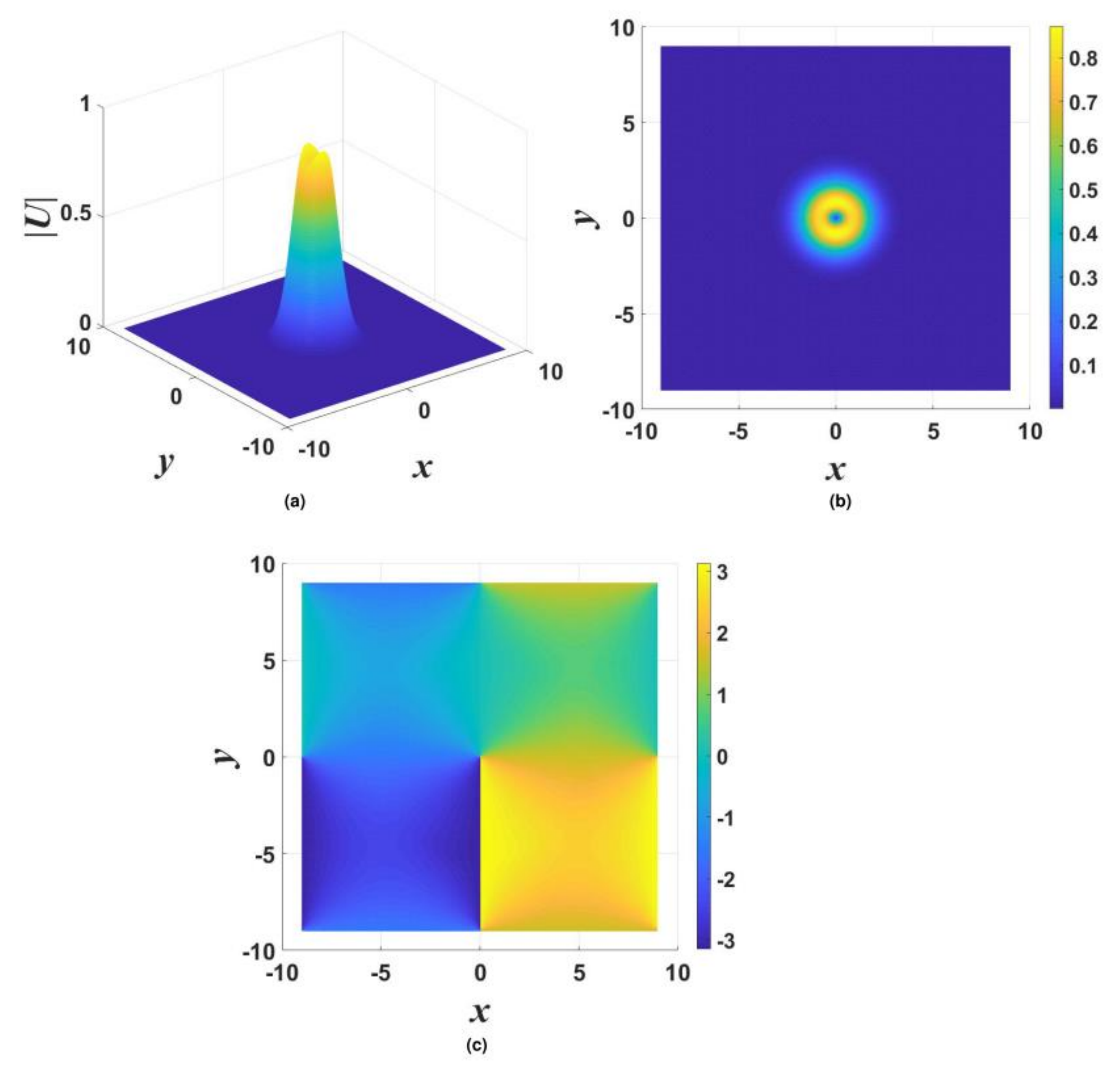}
\caption{The same as in Fig. \protect\ref{fig6}, but for the stable vortex
soliton with $m=1$ in the case of imaginary potential Eq. (\protect\ref{xy}%
), with $\protect\gamma _{0}=0.4$, $\protect\beta =0.5$, $\protect\sigma =0$%
, and propagation constant $k=-3.4$.}
\label{fig13}
\end{figure}

\begin{figure}[tbp]
\centering\includegraphics[width=0.75\textwidth]{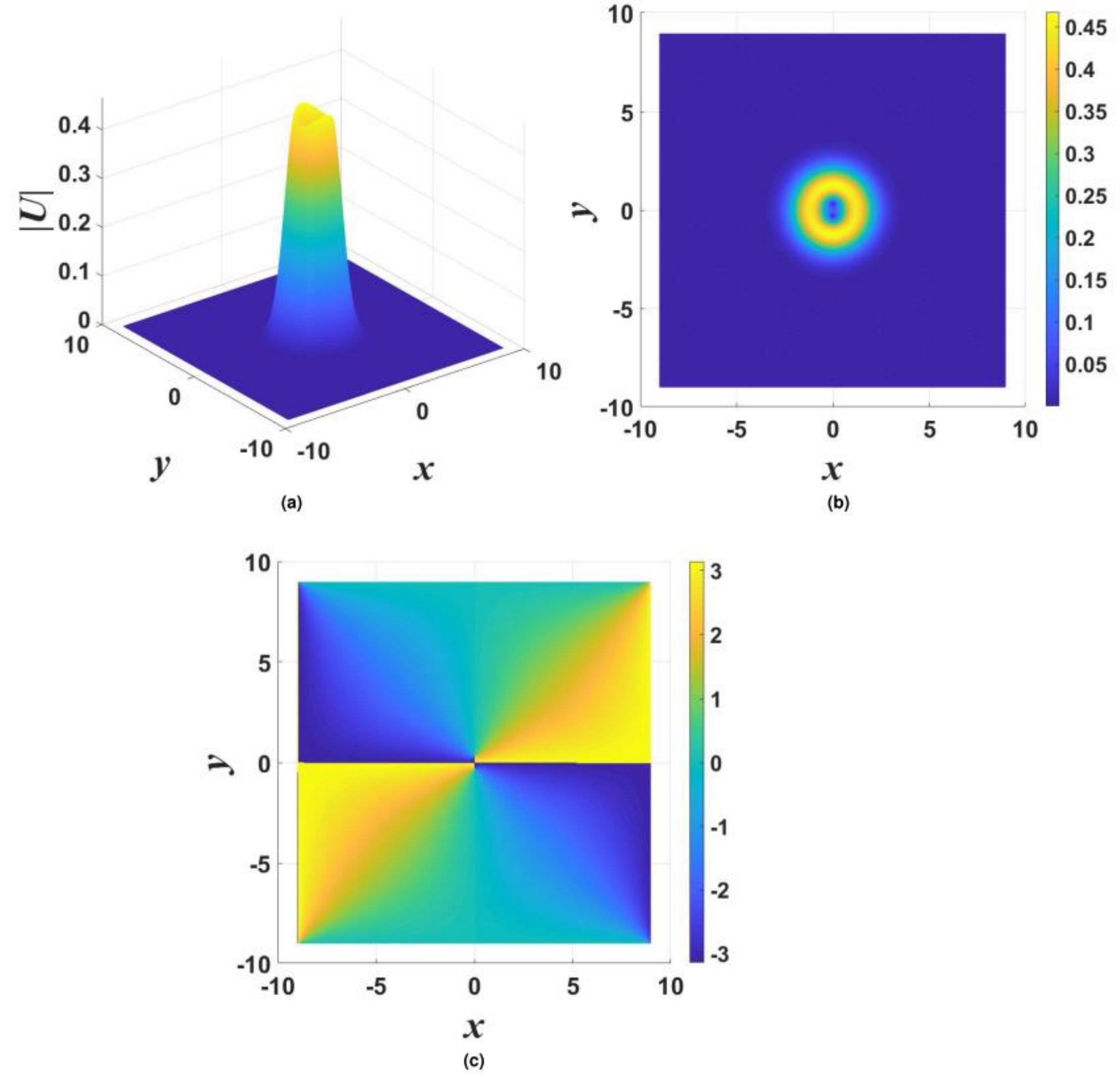}
\caption{The same as in Fig. \protect\ref{fig13}, but for stable vortex
solitons with $m=2$ and $k=-3.6$.}
\label{fig14}
\end{figure}

Starting from input Eq. (\ref{AngularPush}), stable vortices can be
constructed in the model with the gain-loss profile Eq. (\ref{xy}) only if
it is subject to the spatial confinement (recall the same is reported above
for zero-vorticity solitons). Examples of stable solitons with vorticities $%
m=1,2$ and $3$ found in this model are shown in Figs. \ref{fig13} - \ref%
{fig15}. Note that higher-order states with $m\geq 2$ are actually compound
states built of $m$ unitary vortices, whose pivots do not merge into a
single one, remaining separated, although with a small distance between
them, as can be seen for $m=2$ in Fig. \ref{fig14}. The pivots form arrays
along axes $x$ or $y$, the particular direction being randomly chosen by the
initial conditions. Nevertheless, the overall shapes of the unitary and
higher-order vortices are nearly isotropic, due to the structure of the
gain-loss term in Eq. (\ref{xy}) (cf. strongly anisotropic shapes of
vortices in Figs. \ref{fig6}, \ref{fig8}, and \ref{fig9}, supported by the
imaginary potential (\ref{x})).

\begin{figure}[tbp]
\centering\includegraphics[width=0.75\textwidth]{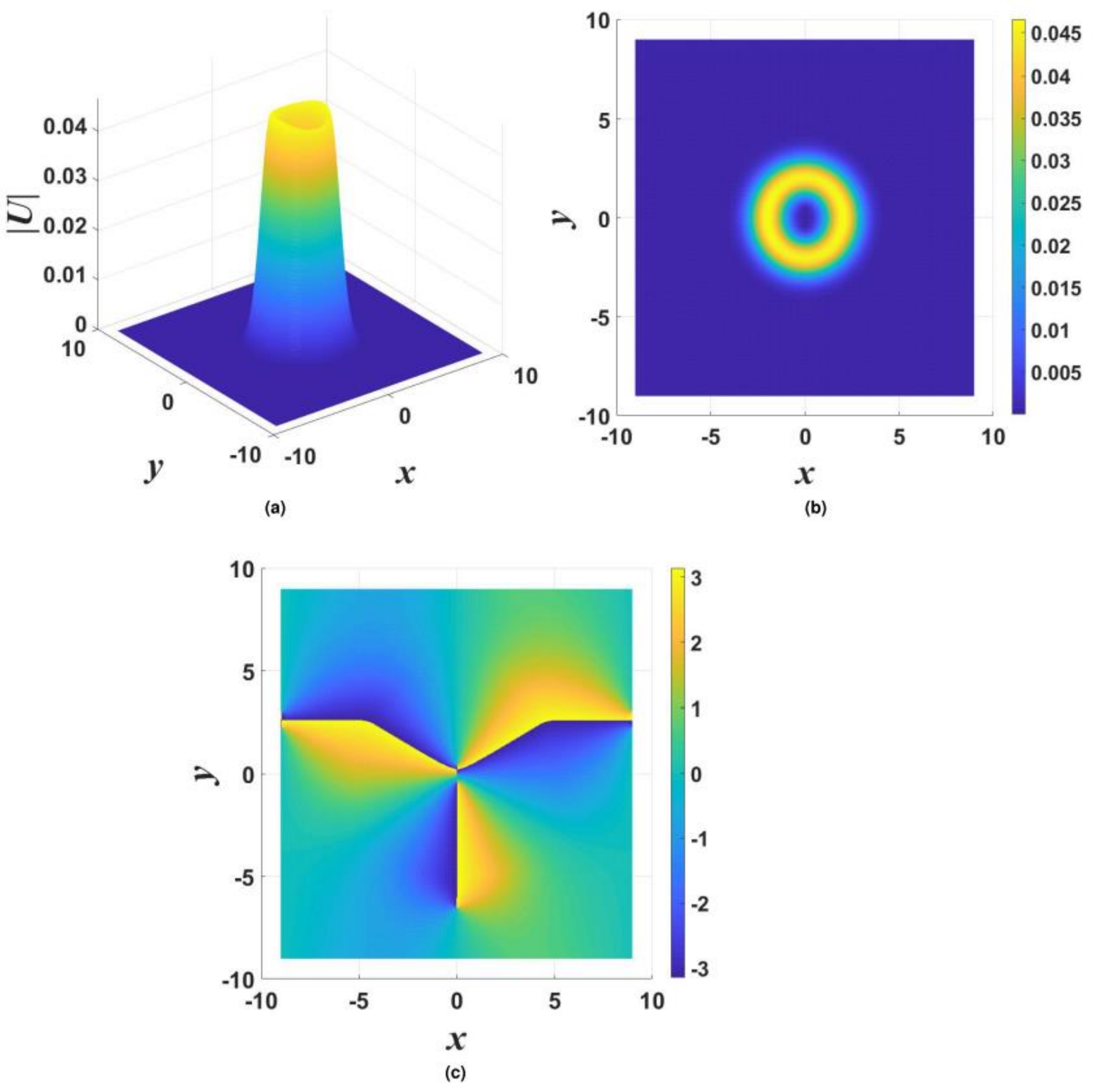}
\caption{The same as in Fig. \protect\ref{fig13}, but for stable vortex
solitons with $m=3$ and parameters $\protect\gamma _{0}=0.2$, $\protect\beta %
=0.5$, $\protect\sigma =0$, $k=-2.2$.}
\label{fig15}
\end{figure}

\begin{figure}[tbp]
\centering\includegraphics[width=0.75\textwidth]{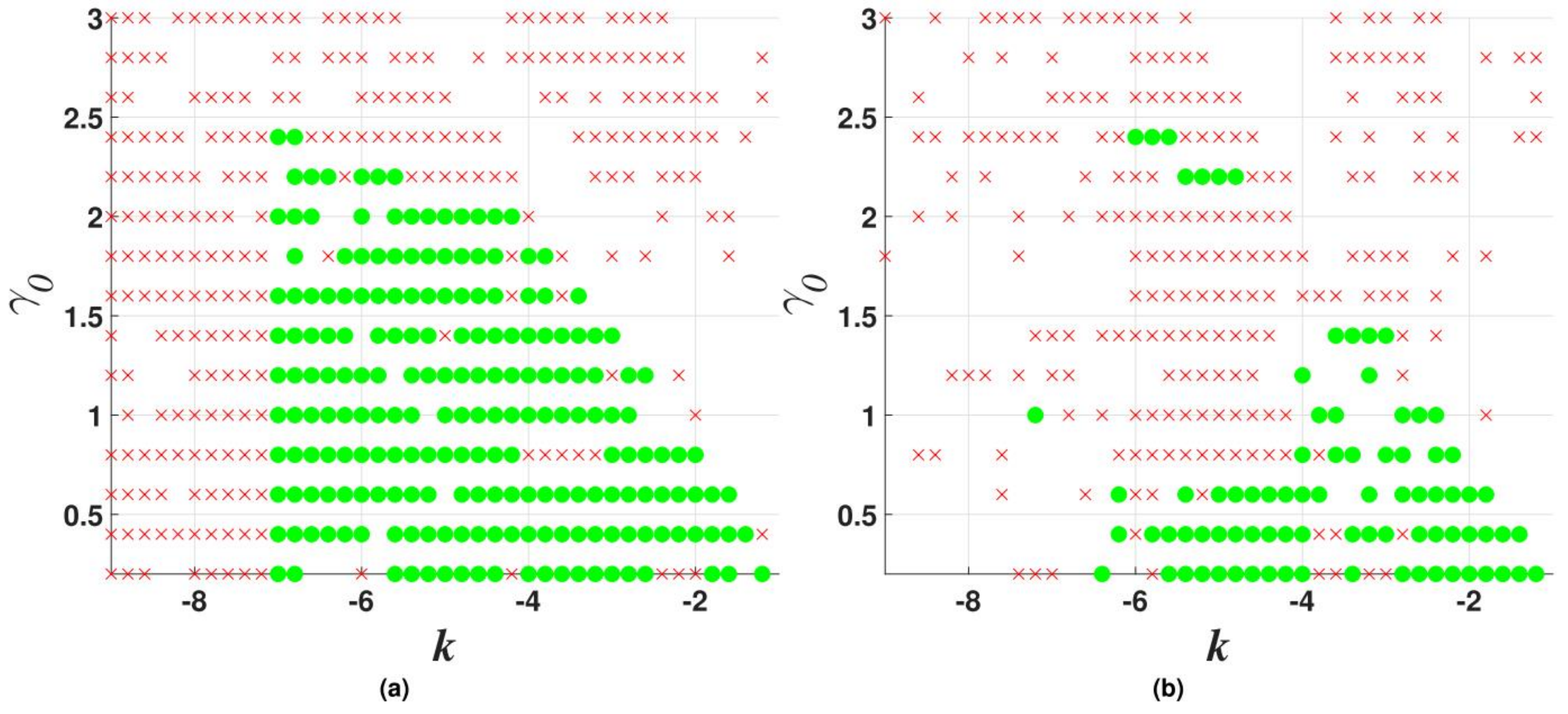}
\caption{Stability charts for vortex solitons with $m=1$ in the model
including imaginary potential (\protect\ref{xy}), with $\protect\beta =0.5$
and $\protect\sigma =0$ in (a) or $\protect\sigma =1$ in (b).}
\label{fig16}
\end{figure}

\begin{figure}[tbp]
\centering\includegraphics[width=0.75\textwidth]{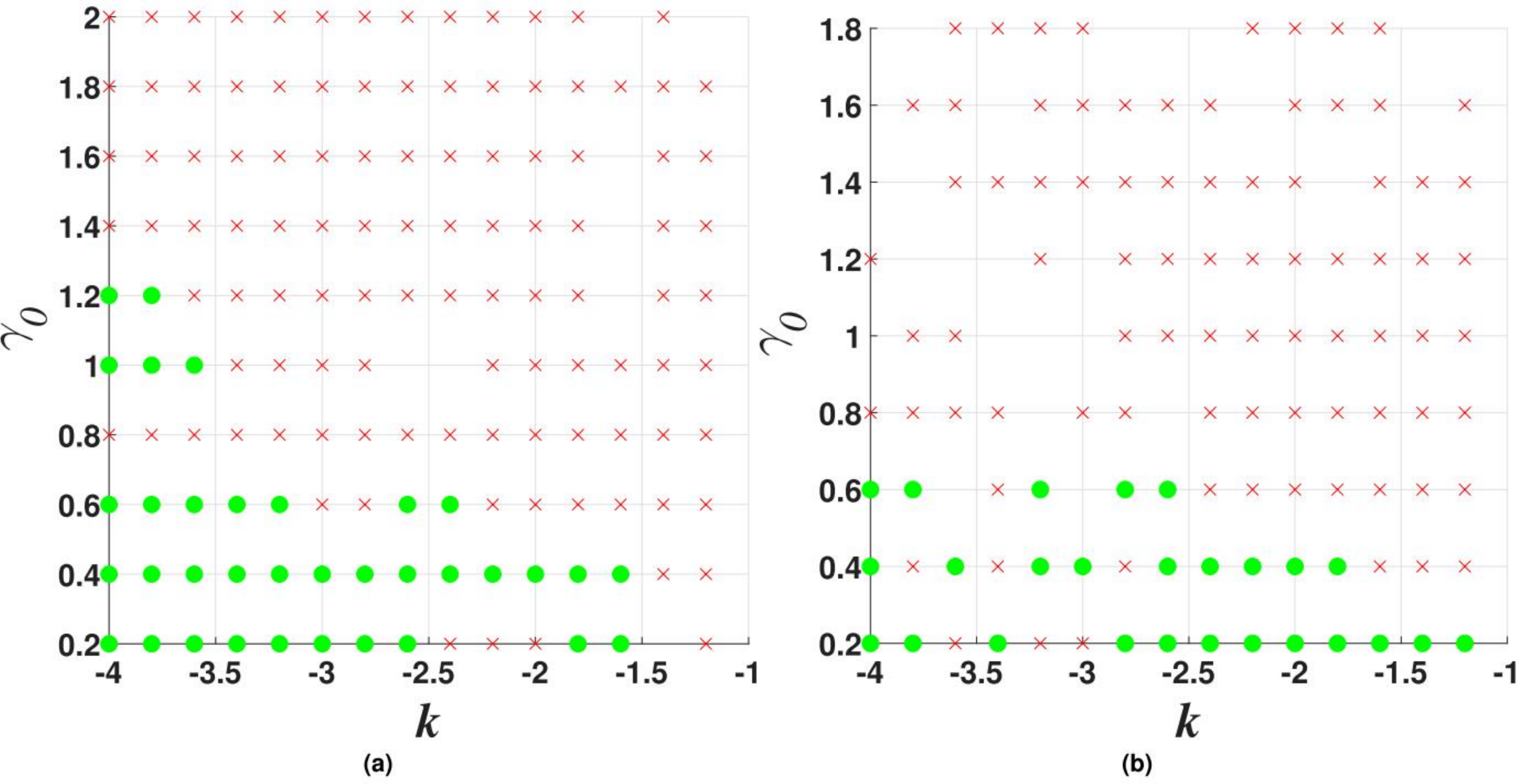}
\caption{The same as in Fig. (\protect\ref{fig16}), but for vorticity $m=2$.}
\label{fig17}
\end{figure}

Stability charts obtained in this model for the solitons with embedded
vorticities $m=1$ and $2$ are shown in Figs. \ref{fig16} and \ref{fig17}.
Only few examples of stable vortices with $m=3$, not shown here, have been
found in this case (for instance, at $\sigma =0$, $\gamma _{0}=0.4$, $k=-1.2$%
).

\begin{figure}[tbp]
\centering\includegraphics[width=0.75\textwidth]{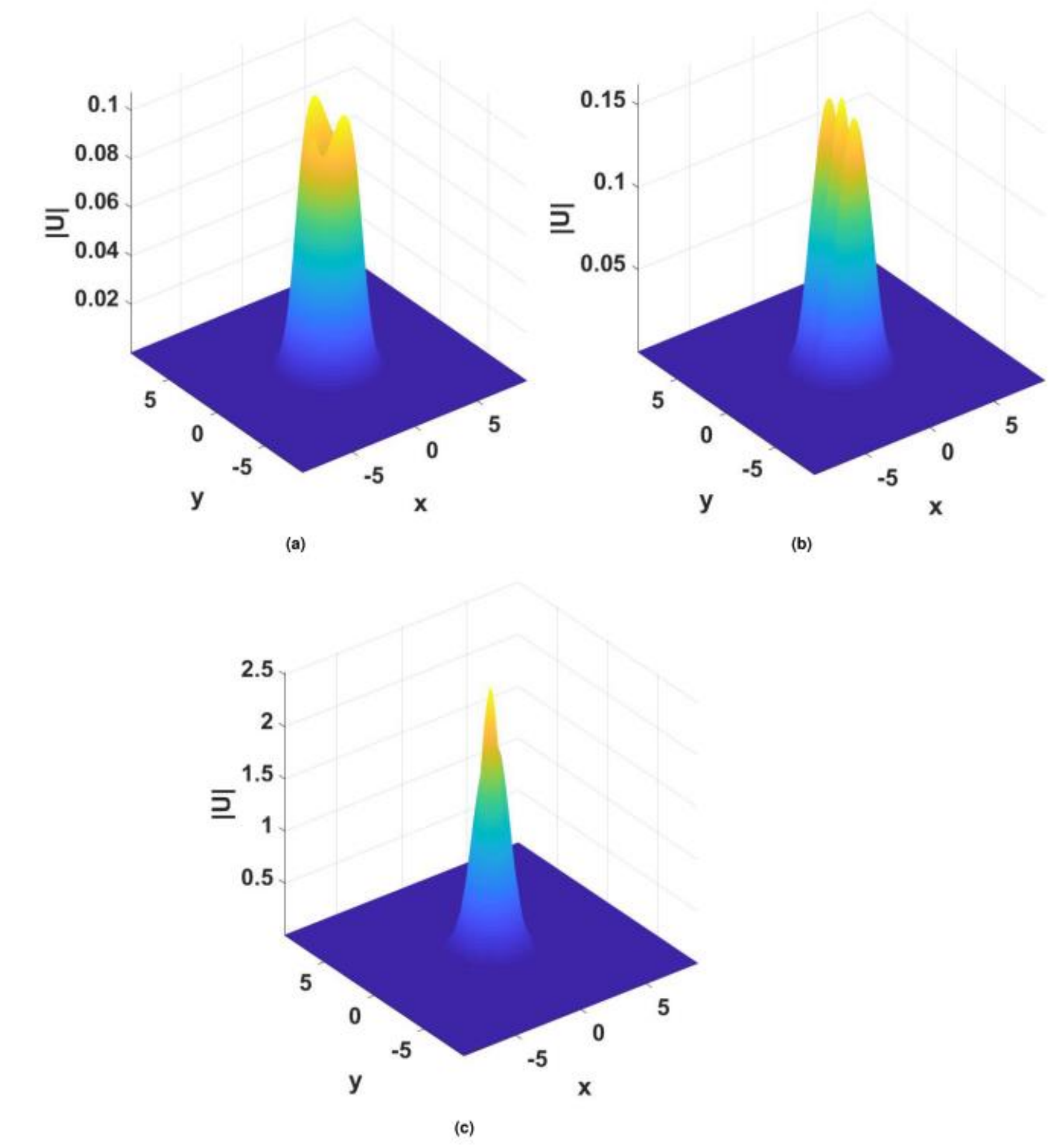}
\caption{The blowup of an unstable vortex soliton with $m=2$ and $\protect%
\gamma _{0}=1.2$, $\protect\beta =0.5$, $\protect\sigma =1$, $k=-2.4$, in
the model with imaginary potential (\protect\ref{xy}). Panels display the
field at $z=60$ (a), $z=200$ (b) and $z=300$ (c). Note the difference in
vertical scales between them.}
\label{fig18}
\end{figure}

Finally, a generic example of the evolution of an unstable vortex soliton is
shown in Fig. \ref{fig18}. The strong difference between vertical scales in
different panels of the figure clearly suggests that the instability leads
to the blowup of the unstable mode, in the course of which the original
vortex tends to fuse into a single peak. In fact, all unstable solitons
considered in this work tend to develop the blowup in direct simulations.

\section*{Discussion}

The objective of this work is to elaborate 2D models with the spatially
modulated self-defocusing nonlinearity and gain-loss distributions
[imaginary potentials, $iW\left( x,y\right) $] which give rise to families
of stable single-peak, multi-peak, and vortical solitons, including ones
which may persist and remain stable (\textquotedblleft unbreakable") at
arbitrarily large values of strengths $\gamma _{0}$ of the imaginary
potential. The unbreakability is possible in the case of the $\mathcal{PT}$%
-symmetric imaginary potential, which is given by Eq. (\ref{x}). An asset of
the models, which can be implemented in bulk nonlinear optical waveguides
with embedded gain and loss elements, is that they produce universal
asymptotic solutions for solitons' tails, along with full exact solutions
for selected species of 2D fundamental and vortex solitons (the latter one
is available in the elliptically deformed version of the model). In
particular, in the limit of large $\gamma _{0}$, the unbreakable family of
fundamental solitons tends to shrink towards the exact solution. Generic
families of zero-vorticity solitons, including single- and multi-peak ones
and higher-order radial states of single-peak solitons, as well as families
of self-trapped modes with embedded vorticity $m=1,2$, and $3$, are
constructed in the numerical form, and their stability is identified by
means of the numerical computation of eigenvalues for small perturbations,
and verified by direct simulations. In the case of the $\mathcal{PT}$%
-symmetric imaginary potential (\ref{x}) the solitons are stable in vast
parameter regions, and feature a trend towards maintaining the unbreakable $%
\mathcal{PT}$ symmetry. Under the action of the imaginary potential (\ref{xy}%
), families of stable fundamental and vortex solitons exist too, provided
that the imaginary potential is subject to spatial confinement.

A relevant extension of the analysis may be to address the elliptically
deformed model, which is considered in the present work in a brief form. A
challenging problem is the possibility of the fractal structure of the
stability patterns in the models' parameter planes.

\section*{Methods}

\subsection*{\textbf{The Newton conjugate gradient method for 2D robust }$PT$%
\textbf{-symmetry model}}

Solutions of the stationary equation (\ref{UU}) were constructed by means of
the Newton conjugate gradient method, which is presented in detail in book
\cite{Yang-book}. In terms of this method, the stationary-solution operator $%
\mathbf{L}_{0}$ is defined by Eq. (\ref{UU}), while the respective
linearization operator $\mathbf{L}_{1}$ is defined as

\begin{equation}
\mathbf{L}_{1}=%
\begin{bmatrix}
A & B \\
C & D%
\end{bmatrix}%
,  \label{L1_linearization_operator}
\end{equation}%
with matrix elements
\begin{eqnarray}
A &=&-k+\frac{1}{2}\nabla ^{2}-\Sigma (r)(\left[ 3(\mathrm{Re}U)^{2}+(%
\mathrm{Im}U)^{2}\right] , \\
B &=&-2\Sigma (r)\mathrm{Re}U\cdot \mathrm{Im}U+W(x,y), \\
C &=&-2\Sigma (r)\mathrm{Re}U\cdot \mathrm{Im}U-W(x,y), \\
D &=&-k+\frac{1}{2}\nabla ^{2}-\Sigma (r)\left[ 3(\mathrm{Im}U)^{2}+(\mathrm{%
Re}U)^{2}\right] ,
\end{eqnarray}%
where the nonlinearity coefficient, $\Sigma (r)$, and imaginary potential, $%
W\left( x,y\right) $ are defined, respectively, by Eq. (\ref{sigma}) and
Eqs. (\ref{x}) or (\ref{xy}).

\subsection*{\textbf{Simulations of the evolution of the wave fields}}

Direct simulations of the evolution equation (\ref{NLS}), written as

\begin{equation}
i\frac{\partial U}{\partial z}=-\frac{1}{2}\left( \frac{\partial ^{2}U}{%
\partial x^{2}}+\frac{\partial ^{2}U}{\partial y^{2}}\right) +\left[
k+\Sigma (r)|U|^{2}+i\gamma \left( x,y\right) \right] U,  \label{EqOnU}
\end{equation}%
cf. Eq. (\ref{UU}), have been performed by means of the commonly known
split-step method. Marching forward in $z$ at each step was split in two
parts, according to the following equations:

\begin{eqnarray}
\mathbf{I} &\mathbf{:~}&i\frac{\partial U}{\partial z}=\left[ k+\Sigma
(r)|U|^{2}+i\gamma \left( x,y\right) \right] U, \\
\mathbf{II} &\mathbf{:~}&i\frac{\partial U}{\partial z}=-\frac{1}{2}\left(
\frac{\partial ^{2}U}{\partial x^{2}}+\frac{\partial ^{2}U}{\partial y^{2}}%
\right) .
\end{eqnarray}

The solutions were numerically constructed in the 2D spatial domain, $%
\left\vert x,y\right\vert \leq 9$, which was covered by a discrete grid of
size $N_{x}\times N_{y}=512\times 512$. The direct simulations were carried
out with step $\Delta z=10^{-5}$. This small step was selected to provide
sufficient accuracy of the numerical solutions obtained in the presence of
the \textquotedblleft exotic" nonlinearity-modulation and gain-loss profiles
(\ref{sigma}) and (\ref{x}) or (\ref{xy}).

\subsection*{\textbf{The stability analysis}}

The stability of the stationary states against small perturbations were
based, as usual, on the general expression for a perturbed solution,%
\begin{equation}
u\left( x,y,z\right) =e^{ikz}\left\{ U\left( x,y\right) +\varepsilon \left[
e^{\Gamma z}v\left( x,y\right) +e^{\Gamma ^{\ast }z}w^{\ast }\left(
x,y\right) \right] \right\} ,  \label{pert}
\end{equation}%
where $\varepsilon $ is an infinitesimal perturbation amplitude, with
eigenmodes $\left\{ v\left( x,y\right) ,w\left( x,y\right) \right\} $ and
(complex) eigenvalue $\Gamma $, which should be found from the numerical
solution of the respective linearized equations,%
\begin{equation}
\begin{array}{c}
\left( -k+i\Gamma \right) v+\frac{1}{2}\left( \frac{\partial ^{2}}{\partial
x^{2}}+\frac{\partial ^{2}}{\partial y^{2}}\right) v-2\Sigma
(r)|U|^{2}v-\Sigma U^{2}w=i\gamma \left( x,y\right) v, \\
\left( -k-i\Gamma \right) w+\frac{1}{2}\left( \frac{\partial ^{2}}{\partial
x^{2}}+\frac{\partial ^{2}}{\partial y^{2}}\right) w-2\Sigma
(r)|U|^{2}w-\Sigma U^{2}v=-i\gamma \left( x,y\right) w,%
\end{array}
\label{eigen}
\end{equation}%
subject to zero boundary conditions at $\left\vert x,y\right\vert
\rightarrow \infty $ (in fact, at borders of the solution domain). These
equations were solved by means of the known spectral collocation method \cite%
{Yang-book}.

\section*{Acknowledgements}

This work is supported, in a part, by the Binational (US-Israel) Science
Foundation through grant No. 2015616, and Israel Science Foundation, grant
No. 1287/17.

\end{document}